\documentclass[11pt]{article}
 \pdfoutput=1

\usepackage{amsmath,amssymb,exscale}
\usepackage{marvosym}
\usepackage{array,multicol}
\usepackage{afterpage,float,flafter}
\usepackage{amssymb}
\usepackage{epsfig,rotating,pifont}
\usepackage{cite}
\usepackage{hyperref}
\usepackage{color}
\usepackage{ marvosym }
\usepackage{textcomp}
\usepackage{wrapfig}

\definecolor{nicered}{rgb}{0.5,0.1,0.1}
\definecolor{nicegreen}{rgb}{0.1,0.5,0.1}
\definecolor{niceblue}{rgb}{0.1,0.1,0.8}
\hypersetup{colorlinks,citecolor= nicegreen,linkcolor= nicered}

\@addtoreset{equation}{section} \makeatother

\def\beq{\begin{eqnarray}}
\def\eeq{\end{eqnarray}}
\def\lsim{\mathrel{\rlap{\lower3pt\hbox{\hskip0pt$\sim$}}
\raise1pt\hbox{$<$}}}         
\def\gsim{\mathrel{\rlap{\lower4pt\hbox{\hskip1pt$\sim$}}
\raise1pt\hbox{$>$}}}         

\newcommand{\ddslash}[1]{#1 \! \! \! \!  {\bf /}}

\title{
\vspace{-3cm}
\begin{flushright}
\small{CERN-PH-TH/154\\
DESY 11-109}
\end{flushright}
\vspace{2.7cm}
\begin{center}
\medskip
 {\Huge\bf Flavor and CP Invariant \\[3mm]
Composite Higgs Models
}
\end{center}
\vspace{0.6cm}
\author{
\Large{\text{\bf Michele Redi$^{1,2}$}\footnote{michele.redi@cern.ch}~~{\Large and} \text{\bf Andreas Weiler$^{1,3}$}\footnote{andreas.weiler@cern.ch}}\\ \\
$^1$\emph{CERN, Theory Division, CH-1211, Geneva 23, Switzerland}\\
$^2$\emph{INFN, 50019 Sesto F., Firenze, Italy}\\
$^3$\emph{DESY, Notkestrasse 85, D-22607 Hamburg, Germany}
}
}

\date{}
\begin{document}
\maketitle \thispagestyle{empty} \vspace*{-.2cm}

\begin{abstract}
The flavor protection in composite Higgs models with partial compositeness is known to be insufficient. We explore 
the possibility to alleviate the tension with CP odd observables by assuming that flavor or CP are symmetries of the composite sector, 
broken by the coupling to Standard Model fields. One realization is that the composite sector has a 
flavor symmetry $SU(3)$ or $SU(3)_U \otimes SU(3)_D$ which allows us to realize Minimal Flavor Violation. We show how to avoid the previously problematic tension between a flavor symmetric composite sector and electro-weak precision tests. Some of the light quarks are substantially or even fully composite with striking signals at the LHC. We discuss the constraints from recent dijet mass measurements and give an outlook on the discovery potential. We also present a different protection mechanism where we separate the generation of flavor hierarchies and the origin of CP violation. This can eliminate or safely reduce unwanted CP violating effects, realizing effectively ``Minimal CP Violation'' and is compatible with a dynamical generation of flavor at low scales.
\end{abstract}

\newpage
\renewcommand{\thepage}{\arabic{page}}
\setcounter{page}{1}

\section{Introduction}

The striking phenomenological success of the Standard Model (SM) flavor sector is potentially threatened by any new physics
that addresses the hierarchy problem. If new degrees of freedom coupled to SM fields appear around the TeV scale, 
as demanded  by naturalness of the electro-weak scale, then the theory has generically more flavor violating structures
than the Cabibbo-Kobayashi-Maskawa (CKM) matrix of the SM and unacceptably large Flavor Changing Neutral Currents 
(FCNC) are generated. 

Two structural solutions to this puzzle may be conceived. Either the flavor structure of new physics is identical
to the SM  or a dynamical mechanism  exists around the TeV scale which approximately aligns new 
physics flavor breaking with the SM. In the first case, which goes under the name of Minimal Flavor Violation (MFV) \cite{MFV1,MFV},
flavor is decoupled from the electro-weak scale and imprinted on the Yukawa matrices at arbitrarily high energies, while in the latter 
flavor signatures can be more revealing and the theory of flavor could be explored at the LHC. 

In this paper we will focus on the flavor problem in Composite Higgs Models (CHM)  
with partial compositeness, where both approaches might be relevant (see \cite{georgikaplan} for earlier work and \cite{RSreview} for reviews). 
Similar considerations can also be applied to Higgless theories \cite{higgsless}. In CHM the Higgs doublet is a composite state, likely a Goldstone Boson, 
of some strongly coupled theory characterized by a dynamical scale $m_\rho$ of a few TeV and a coupling  $1<g_\rho < 4 \pi$. 
The compositeness scale can be larger than in technicolor (or Higgsless) theories which allows in principle to get as close to the SM limit as required. 
Not surprisingly $m_\rho$ can be at most a few TeV if the theory shall remain natural. 
The new ingredient of modern proposals is the idea of partial compositeness. 
The SM fermions and gauge fields are  elementary particles which mix with composite states with equal quantum numbers, 
similarly to the photon-$\rho$ mixing in QCD. The degree of compositeness of fermions increases with their mass so that light 
generations are mostly elementary while the top is strongly composite. Under broad assumptions and despite the presence of 
many new flavor structures, FCNCs are proportional to the elementary-composite mixings and therefore are suppressed for the light generations 
as demanded by observations~\cite{flavorgeneration}. 
This feature offers a largely superior realization of flavor than in classical technicolor theories where fermion masses are generated 
by non-renormalizable interactions. Moreover if the strong sector is a CFT  then the hierarchies of masses and 
mixings can be dynamically generated by the Renormalization Group  Evolution (RGE) of the coupling starting from an anarchic flavor structure 
of the composite sector \cite{minimalcomposite}. Finally, all these ideas can be realized as effective field theories in five dimensions,
\`a la Randall-Sundrum.

Despite these improvements, the flavor protection generated by partial compositeness appears at present insufficient for 
several CP odd observables, most notably in the Kaon system, which indicates a scale of compositeness of 10-20 TeV \cite{weiler}, 
clearly at odds with naturalness. In our opinion this is a strong reason to believe that these models are at best incomplete.
In this paper we will first discuss how the hypothesis of MFV can be realized in CHM. This erases the flavor problem at the price
of abandoning the dynamical generation of flavor. The possibilities can be divided in two categories. In the approach pioneered 
in \cite{rattazzizaffaroni} all the SM fermions are composite states and the SM flavor structure is directly shined on the Yukawas of the 
composite sector through appropriate flavon fields. Clearly full compositeness is severely constrained by a variety 
of precision measurements.  Alternatively, and we will focus on this possibility here, the flavor structure may arise from the 
mixings if the strong sector is flavor symmetric. This implies that the composite sector has global flavor symmetries, like QCD.
This idea has been considered in the past within five-dimensional constructions \cite{Cacciapaglia:2007fw}. 
We extend this setup to general CHM within an effective four dimensional description and present novel scenarios where the flavor structure originates 
from the mixings of left-handed fields. One important ingredient will be the mixing into split left-handed doublets, 
where each left-handed doublet couples to two different states of the composite sector.
Realizing MFV demands some of the light quarks to be significantly composite.  We show that constraints from precision electro-weak 
measurements can be avoided. As a bonus we show that Electric Dipole Moments (EDM) are also suppressed compared to the standard scenario, 
a feature not granted by MFV.  In particular we find that the new scenario with composite right-handed  quarks is very weakly 
constrained by present data, allowing  these chiralities to be strongly or even entirely composite. In this case very exciting 
phenomenology is expected at the LHC where large cross-sections for the production of new resonances are expected.

Within the same framework we are also led to a different solution which hinges
on the assumption that the composite sector is CP invariant. One intriguing feature is that most of the experimental tension 
originates from CP odd observables. One natural idea is to try to separate the origin of CP violation in the SM from the origin of the flavor hierarchies. 
This is impossible within the SM where only the CKM matrix is physical but it might be possible in CHM where more flavor 
structures exist. If the strong sector respects CP, then this symmetry must be broken by the mixings in order to reproduce the $\mathcal{O}(1)$ CP 
phase of the CKM matrix. Depending on how CP violation is communicated to the SM very different effects are obtained, not necessarily
improving the phenomenology (except for EDMs).  We find however that if CP violation originates from the mixing of third generation quarks, then all CP 
violating effects beyond the SM  are automatically suppressed for the light generations, allowing us to evade all
experimental constraints with a low scale of compositeness. We dub this mechanism ``Minimal CP Violation''.
Interestingly, this is compatible with the dynamical generation of flavor structure at low scales.

This paper is organized as follows: In section \ref{model} we introduce the CHM that we will study in the language of a simple
two site model which includes only the first resonances of the composite sector. We discuss the relevant 
representations of the composite states with emphasis on the case of split left-handed fermions 
which has special phenomenological features and we review the status of experimental bounds in these models.
In section \ref{MFV} we discuss the different possibilities to realize MFV assuming a flavor symmetric strong sector.
Two possibilities stand out with composite left-handed or right-handed quarks. We study the experimental bounds from precision  
measurements and compositeness limits. We outline the  phenomenology of this model at the LHC 
where cross-section much larger that in the standard CHM could be obtained.
In section \ref{CPprotection} we study the possibility that the strong sector is CP invariant. This allows us to separate
the origin of flavor and CP violation. We show that if CP violation is induced from the mixing of third generation quarks,
unwanted CP violating effects can be strongly suppressed. We end in section \ref{outlook}.

\section{Framework}
\label{model}

The physics of composite Higgs models that we wish to study 
can be efficiently captured in terms of a 2 site model where the composite sector is replaced by
the first resonances which mix with the SM fields\cite{2sitescontino}\footnote{This picture should be refined in the case  where the Higgs is pseudo-Goldstone boson~\cite{discrete,4dcomposite}. However one obtains similar results in both  cases so we will use the language of the two site lagrangian in the rest of the paper.}. 
Most minimally, besides the Higgs doublet, the strong sector contains massive Dirac fermions in the same representation 
of the SM fermions under the SM gauge group as well as spin 1 resonances of SM gauge fields. The lagrangian reads \cite{2sitescontino}\footnote{For a related discussion  see also Ref. \cite{buras}. We focus on the quark sector, the extension to leptons being trivial in what follows.},
\begin{eqnarray}
{\cal L}_{composite}&=& -\frac 1 4 \rho_{\mu\nu}^{i2} + \frac {m_\rho^{i2}} 2 \rho_\mu^{i2}\nonumber + |D_\mu H|^2- V(H)\\
&&\bar{Q}^i (i \ddslash{ D} -  m_Q^i) Q^i + \bar{U}^i (i \ddslash{ D} -  m_U^i) U^i + \bar{D}^i (i \ddslash{ D} -  m_D^i) D^i \nonumber \\
&&+Y^U_{ij} \bar{Q}_L^i \tilde{H} U_R^j+ Y^D_{ij} \bar{Q}_L^i H D_L^j
+\tilde{Y}^U_{ij} \bar{Q}_R^i \tilde{H} U_L^j+ \tilde{Y}^D_{ij} \bar{Q}_R^i H D_L^j+ {\rm h.c.}
\label{2site}
\end{eqnarray}
where $\rho_\mu^i$ are massive $SU(3)_c\otimes SU(2)_L \otimes U(1)_Y$ gauge fields coupled to the fermions through 
standard covariant derivatives. By assumption the strong sector does not have large hierarchies and can be characterized by a
single mass scale $m_\rho$  and a coupling $g_\rho$. We will however keep the Yukawa couplings and gauge couplings 
independent since even small hierarchies between them can produce important phenomenological effects.

The elementary sector includes chiral fermions with SM quantum numbers and massless $SU(3)\otimes SU(2)\otimes U(1)$ gauge fields. Importantly, SM fields are coupled linearly to operators of the strong sector, inducing a mixing between the states of the two sectors
\begin{equation}
{\cal L}_{mixing}=m_\rho \left[\lambda_q^{ij} \bar{q}_{Li} Q_{Rj} + \lambda_u^{ij} u_{Ri}   \bar{U}_{Lj}+ \lambda_d^{ij} d_{Ri}   \bar{D}_{Lj}+{\rm h.c.}\right]
\end{equation}
Due to these mixings the SM quarks are a rotation of elementary and composite fermions.
The mixings are in general complex matrices which can be diagonalized via a bi-unitary transformation,
\begin{equation}
\lambda= L\cdot \hat{\lambda}\cdot R
\label{lambdadecomposition}
\end{equation}
The left matrix is not physical and can be eliminated with a redefinition of the elementary fields. In this basis,
the mass eigenstates of left doublets (before electro-weak symmetry breaking)  are given by,
\begin{equation}
\left(\begin{array}{c}
q_{Li} \\
Q_{Li}
\end{array}\right)=\left(\begin{array}{cc}
\cos \varphi_{q_{Li}} & - \sin \varphi_{q_{Li}} \\
 \sin \varphi_{q_{Li}} & \cos \varphi_{q_{Li}}
\end{array}\right)
\left(\begin{array}{c}
 q^{el}_{Li} \\
R^L_{ij} Q^{co}_{Lj}
\end{array}\right)
\label{rotationleftdoublets}
\end{equation}
and similarly for the singlets. Note that we have included a unitary matrix needed to connect the two bases in general. 
Rotating to the mass basis one obtains the SM Yukawa couplings,
\begin{eqnarray}
y_{ij}^u&= f_{q_{Li}} \left(Y^U\right)_{ij} f_{u_{Rj}}^\dagger\nonumber \\
y_{ij}^d&= f_{q_{Li}} \left(Y^D\right)_{ij} f_{d_{Rj}}^\dagger
\label{yukawas}
\end{eqnarray}
where the mixing functions are given by,
\begin{equation}
f_{\psi_i}=\sum_{j=1}^3 \sin\varphi_{\psi_j} R^\psi_{ji}\qquad \psi = q_L,u_R,d_R
\label{mixingfunctions}
\end{equation}
In the standard construction the SM Yukawa hierarchies are generated by hierarchies of the mixing angles
for which a dynamical origin can be realized. 
If $Y_{u,d}$  are general complex matrices the matrices $R^\psi$ are redundant and can be dropped. 

Similarly the SM gauge fields mix with the heavy spin-1 resonances in a way entirely analogous to the 
photon-$\rho$ mixing in QCD. Denoting the coupling of the elementary gauge fields $g_{el}$, upon diagonalization one has,
\begin{equation}
g= g_{el} \cos\theta\,,~~~~~~~~~~~~~~~~~~~~~~
\tan \theta = \frac{g_{el}} {g_\rho}.
\end{equation}
where $g$ is the SM gauge coupling.
 
\subsection{Choice of Representations}
\label{choicerep}

The basic picture reviewed above must be refined to build realistic scenarios where the Higgs is a composite state.
Most importantly the strong sector which delivers the Higgs should respect the custodial symmetry $SU(2)_L\otimes SU(2)_R$ needed to protect the $T$ parameter from large tree level corrections. This is an accidental symmetry of the renormalizable SM lagrangian which needs to be promoted to a true symmetry of the strong sector in order to avoid isospin violations from dimension 6 operators suppressed by a low compositeness scale. To reproduce the hyper-charge  of SM fermions the composite states should also carry  a $U(1)_X$ charge so that,
\begin{equation}
Y=T_{3R}+X
\end{equation}
When the Higgs is pseudo-Goldstone boson arising from the breaking $G/H$, with $SU(2)_L\otimes SU(2)_R\in H$, there can also be 
states associated to $G$ representation of the strong sector. We omit these as they will not play an important role in what follows.

Composite states are classified according to representations of $SU(2)_L\otimes SU(2)_R\otimes U(1)_X$.
The choice of left SM fermions coupled to states in the $(2,1)_{\frac 1 6}$ representation is phenomenologically strongly disfavored. The argument runs as follows \cite{silh}. By custodial symmetry the right-handed SM fermions should couple to fermions in the $(1,2)_{\frac 1 6}$ rep. This mixing breaks custodial symmetry and leads to a sizable loop correction to the $T$ parameter. The leading contribution due to the top right can be roughly be estimated as,
\begin{equation}
\alpha_{em} T\sim \frac 3 {16\pi^2} Y_U^4 \sin^4 \varphi_{t_R} \frac {v^2}{m_\rho^2}.
\end{equation}
The tree level correction to the coupling of the $Z$ to fermions can be estimated in the two site model as~\cite{2sitescontino},
\begin{equation}
\delta g \approx \frac {Y_{U,D}^2 v^2}{2\,m_\rho^2}\sin^2 \varphi_{\psi}\left(T'_{3L}(Q)-T_{3L}\right)+g_{*2}^2\, \frac {v^2}{4 m_\rho^2} \sin^2\varphi_{\psi}(T_{3R}-T_{3L})
\label{deltagl}
\end{equation} 
where the first contribution is due to the mixing, after electroweak-symmetry 
breaking, of the SM fermions with a heavy fermion $Q$ of different electro-weak charge, 
while the second is generated by the mixing of the $Z$ with vector resonances. 
The contribution to $T$ grows with the mixing of $t_R$ while the correction to $g_{b_L}$ increases with the mixing of the left
doublet. Since the product of the two mixings is constrained by the top mass, both bounds cannot be satisfied simultaneously,
if the scale of compositeness is around 3 TeV as required by the $S-$parameter.

A better choice is to couple the left fermions to states in the $(2,2)$ representation of custodial symmetry (the 4 of $SO(4)$), 
as in this case the right-handed top can be coupled to a singlet of custodial symmetry and does not contribute to the $T$-parameter. 
The $(2,2)_{\frac 2 3}$ fermions can be represented as the matrix,
\begin{equation}
L_U=\left(\begin{array}{cc}
T & T_{\frac 5 3} \\
B & T_{\frac 2 3}
\end{array}\right)
\label{rep1}
\end{equation}
where $(T,~B)$ has the same quantum numbers as the SM left doublet and $(T_{5/3},~T_{2/3})$ is an exotic doublet of hypercharge 7/6. 
Yukawas for the up sector are generated by coupling the up right-handed quarks to fermions in the $1_{\frac 2 3}$ rep. The correction to SM couplings of left-handed fermions
reads \cite{protectionzbb},
\begin{equation}
\frac g {\cos \theta}\left[\frac {c_2-c_1} 2\, \bar{b}_L \gamma^\mu b_L- \frac {c_1+c_2+2 c_3} 2\, \bar{t}_L \gamma^\mu t_L\right]\,Z_\mu-\frac g {\sqrt{2}} (c_2+c_3)\bar{t}_L \gamma^\mu b_L W_\mu^\dagger +h.c.
\label{operatorzbb}
\end{equation}
where $c_1$, $c_2$ and $c_3$ are coefficient of effective operators generated by integrating out the strong dynamics. 
If the strong sector possesses a discrete symmetry $L\leftrightarrow R$, then $c_2=c_1$ so that the coupling 
of the $b_L$ is not renormalized (at zero momentum) while the correction of the other two couplings 
are related. From eq. (\ref{deltagl}) we estimate,
\begin{equation}
\delta g_{t\bar{t}}=\delta g_{t\bar{b}} \approx \frac {Y_U^2 v^2}{2\,m_\rho^2}\sin^2 \varphi_{q_L}\
\label{deltaglest}
\end{equation} 
 
We note that, when the Higgs is a Goldstone boson  of the strong dynamics, even if the strong sector is not $Z_2$ symmetric, 
this symmetry  often arises as an accidental symmetry of the two derivate effective lagrangian~\cite{cthdm}.
This is due to the fact that the spontaneously broken symmetry $G$ imposes additional selection rules
which might forbid the generation of couplings that violate the $L\leftrightarrow R$ symmetry. We will assume in what follows 
that the $Z_2$ protection, either exact or accidental is realized in our models.
 
In order to generate the Yukawa couplings for the down quarks, one possibility is that the latter are coupled to fermions
in the $(1,3)_{\frac 2 3}$ representation since a singlet is forbidden by $U(1)_X$ symmetry. 
This can be done in particular respecting the $Z_2$ symmetry above so that the correction to the coupling 
of $b_L$ (or any  left-handed down quarks) to the $Z$ is zero at zero momentum. 

In this paper we will focus on the alternative  possibility where the left doublet couples to two distinct operators,
one responsible for the generation of up Yukawas and the other for down Yukawas. This choice is also 
quite natural since it allows to treat the up and down quarks symmetrically. Assuming that $d_R$ 
couples to a $1_{-\frac 1 3}$ state, Yukawas for the composite fermions are allowed if $q_L$ also couples to a $(2,2)_{-\frac 1 3}$ composite fermion.
The embedding is in this case,
\begin{equation}
L_D=\left(\begin{array}{cc}
B_{-\frac 1 3 } & T'  \\
B_{-\frac 4 3 } & B'
\end{array}\right)
\end{equation}
The mixing breaks the $L\leftrightarrow R$ symmetry for $b_L$ and produces a shift on the coupling of the $Z$ while the coupling of 
$t_L$ is protected. One important consequence of having two different operators for up and down quarks is that, since the size of the mixings can be different, the correction to the coupling of $b_L$ is suppressed if $\lambda_{b_L}\ll \lambda_{t_L}$\footnote{In models with a conformal sector which  explain the generation of flavor hierarchies through RGE it was shown that a hierarchy between the two left mixings is naturally generated \cite{custodians}. In the present scenario realizing MFV, the RGE can not be used to generate the full flavor structure but only a hierarchy between up and down quarks.}. Such a hierarchy is also welcome to reproduce the ratio $m_b/m_t$ and we will work under this assumption in what follows.

The situation is simpler for the right-handed quarks which are coupled to singlets of the strong sector carrying only $U(1)_X$ charge. 
Repeating the operators analysis of \cite{protectionzbb} it is easy to see
that no correction to the coupling of the $Z$ is generated, independently of whether the strong sector is $Z_2$ symmetric or not. 
This can also be checked from equation (\ref{deltagl}). The second term is obviously zero because the states are singlets, while the first term 
is zero, say for the up quarks, due to the cancellation of the shifts from the mixing of $t_R$ with $T$ and $T_{2/3}$ which have opposite $T_{3L}$.

To summarize our setup the right-handed quarks are coupled to singlets of the custodial symmetry with $X$ charge equal to the electric charge
while the left doublets are coupled to two different operators in the $(2,2)$ representation of custodial with $X=2/3,-1/3$, 
one responsible for the generation of Yukawas of the up sector and the other for the Yukawas of the down sector. 
While natural, this choice  is particularly favorable phenomenologically because only the couplings 
of left-handed up quarks to SM gauge bosons are modified in a significant way. 
For our estimates we will use the following ``improved'' effective lagrangian
which captures the relevant phenomenological features
\begin{eqnarray}
{\cal L}_{strong}&=& -\frac 1 4 \rho_{\mu\nu}^{i2} + \frac {m_\rho^{i2}} 2 \rho_\mu^{i2}\nonumber + Tr[|D_\mu H|^2]- V(H)\nonumber \\
&+&Tr[\bar{L}_U^i (i \ddslash{ D}-  m_{L_U}^i) L_U^i] +Tr[\bar{L}_D^i (i \ddslash{ D} -  m_{L_D}^i) L_D^i] +\bar{U}^i (i \ddslash{ D} -  m_U^i) U^i + \bar{D}^i (i \ddslash{ D}-  m_D^i) D^i\nonumber \\
&+&Y^U_{ij} Tr[\bar{L}_{U}^i  {\cal H}]_L U_R^j+ Y^D_{ij} Tr[\bar{L}_{D}^i {\cal H}]_L D_R^j+\tilde{Y}^U_{ij} Tr[\bar{L}_{U}^i {\cal H}]_R U_L^j+ \tilde{Y}^D_{ij} Tr[\bar{L}_{D}^i {\cal H}]_R D_L^j+h.c.\nonumber \\
\label{2sitecustodial}
\end{eqnarray}
where we use matrix notation for the Higgs with ${\cal H} = (\tilde{H}, H)$ and $\tilde{H} = i \sigma_2 H^*$, and where $\rho_\mu^i$ are $SU(3)_c\otimes SU(2)_L\otimes SU(2)_R\otimes U(1)_X$ gauge fields. The mixing terms are,
\begin{equation}
{\cal L}_{mixing}=m_\rho[\lambda_{Lu}^{ij} \bar{q}_{L}^i Q_{Ru}^j+ \lambda_{Ld}^{ij} \bar{q}_{L}^i Q_{Rd}^j + \lambda_{Ru}^{ij} u_{R}^i   \bar{U}_{L}^j+ \lambda_{Rd}^{ij} d_{R}^i   \bar{D}_{L}^j]
\label{mixingcustodial}
\end{equation}
where $Q_{Ru}$ is the $(T,B)_R$ fragment inside $L_u$ and $Q_{Rd}$ the $(T',B')_R$ fragment in $L_d$.
The Yukawas are now,
\begin{eqnarray}
y^u&= f_{Lu} \cdot \left(Y^U\right)\cdot f_{Ru}^\dagger\nonumber \\
y^d&= f_{Ld} \cdot \left(Y^D\right)\cdot f_{Rd}^\dagger
\label{yukawascustodial}
\end{eqnarray}
For notational simplicity we have promoted  the mixing functions to diagonal matrices.

\subsection{CP Violation}
\label{cpreview}

For a compositeness scale of 3 TeV, which is compatible with electro-weak precision tests (in particular the S parameter),
the main bound on CHM originates from CP odd observables. In this section we review the experimental status of CP 
violation in CHM, with emphasis on the choice of representations discussed above which possess special phenomenological features.

\subsubsection{Flavor Changing Processes}

If the Yukawas of the composite sector are anarchic complex matrices the theory contains many new flavor 
violating structures and CP violating phases, diagonal and off-diagonal. Remarkably partial compositeness automatically reduces 
the flavor violating effects associated to the light generations because these are suppressed by the mixing. 
It is easy to see
that 4-Fermi operators are generated with the following parametric suppression,
\begin{equation}
\sim f_q^i  f_q^{\dagger j}  f_q^k  f_q^{\dagger l}\, \frac {g_\rho^2}{m_\rho^2}\, \left(\bar{q}^i q^j \bar{q}^k q^l\right)
\label{extFCNC}
\end{equation}
in the basis (\ref{yukawascustodial}).  For example, the Wilson coefficient $C_4$ of 
LR 4-Fermi operators $\bar{q}_R^{i\alpha} q_{L\alpha}^j \bar{q}_{L}^{i\beta} q_{R\beta}^j$ is suppressed by \cite{weiler},
\begin{equation}
\sim \frac  1 {m_\rho^2}\frac {g_\rho^2}{Y_{U,D}^2} \frac {2 m_i m_j}{v^2}
\label{LRoperator}
\end{equation}

Note, that this estimate does not hold in general when $q_L$ mixes to two different fermions, as in our model.  
If $q_L$ couples to a single state as in the standard case, using the singular value decomposition of the mixing matrix, there is only one unitary matrix on the left which is unphysical and can be removed by a field redefinition of $q_L$ as we did in eq. (\ref{rotationleftdoublets}).
It is then possible to prove that the matrices needed to diagonalize the Yukawas have hierarchies similar to the CKM matrix and the quark masses. This is  crucial to show that the flavor suppression (\ref{extFCNC}) survives after the rotation to the quark mass basis. 
When left-handed fermions couple to two different states the relative misalignment of the  left mixings is physical 
and it is not related in general to the hierarchies of masses and mixings. 
The rotation matrices are not hierarchical unless $D_L$  and $U_L$ are aligned and one can obtain, after rotation to the physical basis, 
larger flavor violations from contributions associated to the third generation in eq. (\ref{extFCNC}).
This can be avoided for example by imposing UV flavor symmetries that align the matrices $D_L$ and $U_L$ \cite{Csaki:2008eh}.
In this case the relevant left mixing in eq. (\ref{extFCNC}) is the larger one.

With the above caveat, the mechanism of partial compositeness works quite well for the effective operators
in the $B$ and $D$ systems  but is in general insufficient for CP violation in the Kaon system. 
Experimentally one finds $\rm{Im}[C_{4K}]<3.9 \times 10^{-11}$ TeV$^{-2}$~\cite{utfit}. Using the above estimate we derive~\cite{weiler,kkbar},
\begin{equation}
m_\rho \gsim 10\, \frac { g_{\rho}}{Y^D}\, \rm{TeV}
\label{epsilonkbound}
\end{equation}

\begin{figure}[t]
\begin{center}
\includegraphics[scale=0.75]{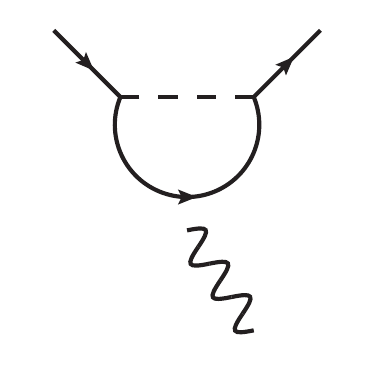}
\caption{Penguin diagram. The photon attaches to the fermion and to charged scalars.}
\label{fig:penguin}
\end{center}
\end{figure}

These are by no means the only problematic observables. Loops of composite dynamics generate dipole operators,
\begin{equation}
	Q_7 =  \frac{e\,m_b}{8 \pi^2}\, \bar b \sigma^{\mu\nu} F_{\mu\nu} (1 - \gamma_5) s \qquad
	Q_7' =  \frac{e\,m_b}{8 \pi^2}\, \bar b \sigma^{\mu\nu} F_{\mu\nu} (1 + \gamma_5) s
\end{equation}
which contribute to the $b\to s \gamma$ transition (both to the real and imaginary parts of the amplitude). 
These are generated by diagrams as in Fig. \ref{fig:penguin}.
with composite fermions and Higgs, $W$ or $Z$ bosons propagating in the loop.
Using the results of \cite{2siteskaustubh, Delaunay:2012cz,straub} we estimate the leading contribution to the Wilson coefficients 
of the operators above as,
\begin{eqnarray}
C_7(m_\rho)&\sim\displaystyle{\frac {v}{8\, m_\rho^2 G_F m_b V_{ts}}} \left[D_L^\dagger \cdot f_{Ld}\cdot Y^D \cdot \tilde{Y}^{D\dagger}\cdot Y^D\cdot f_{Rd}^\dagger \cdot D_R\right]_{23} \nonumber \\
C_7'(m_\rho)&\sim\displaystyle{\frac {v}{8\, m_\rho^2 G_F m_b V_{ts}}} \left[D_L^\dagger \cdot f_{Ld}\cdot Y^D \cdot \tilde{Y}^{D\dagger}\cdot Y^D\cdot f_{Rd}^\dagger \cdot D_R\right]_{32} 
\label{bsgamma}
\end{eqnarray}
where $C_7$ and $C_7'$ are defined by ${\cal H}_{eff} (b\to s \gamma)=-G_F/\sqrt{2} V_{ts}^* V_{tb}[ C_7(\mu_b)Q_7+ C_7'(\mu_b)Q_7']$.

Assuming similar size for $Y$ and $\tilde{Y}$ from eq. (\ref{bsgamma}) one finds,
\begin{equation}
C_7\sim \frac 1 {8}  \left(\frac {Y^D}{m_\rho}\right)^2 \frac {\sqrt{2}}{G_F}\,,~~~~~~~~~~~~~~~~C_7'\sim \frac 1 {8} \left(\frac {Y^D}{m_\rho}\right)^2 \frac {\sqrt{2}}{G_F}\frac {m_s}{m_b V_{ts}^2}
\end{equation}
from which it follows the bound,
\begin{equation}
m_\rho\gsim  0.7 \, Y^D\,\rm{TeV}.
\label{bsgammabound}
\end{equation}
These estimates are then compatible with experimental bounds unless very large values of the coupling are considered.

The contribution to direct CP violation in $\epsilon'/\epsilon_K$ has a spurion dependence similar to $b\to s \gamma$.
In this case one derives a more stringent bound \cite{Gedalia:2009ws},
\begin{equation}
m_\rho\gsim  1.5 \, Y^D \,\rm{TeV}
\label{epsilonprimebound}
\end{equation}
Dipole bounds scale inversely to the one in eq. (\ref{epsilonkbound}) so that both
cannot be satisfied invoking Yukawa couplings larger than the spin 1 coupling $g_\rho$.

\subsubsection{EDMs}
\label{edmsec}

Among flavor diagonal observables the strongest bounds arise from electric dipole moments.
Experimentally the EDM of the neutron is bounded by~\cite{Baker:2006ts}
\begin{equation}
|d_n| < 2.9 \times 10^{-26}\, \rm{e\,cm}\,,~~~~~~~~~~~~~~~~~~( 90\%\, {\rm C.L.})
\end{equation}
In composite Higgs models  contributions to EDMs of up and down quarks arise from the same type of diagrams that contribute to $b\to s \gamma$,
with different external legs. These contributions were estimated in \cite{soni} in the context of Randall-Sundrum scenarios and we here derive
similar estimates in 2-site model. For the down quark we obtain the contribution\footnote{We should in principle consider
the running of this operator from the scale $m_\rho$ where it is generated to the low scale. This is to good approximation taken into account by evaluating
the quark masses at the matching scale 3 TeV.},
\begin{eqnarray}
d_d \sim   \displaystyle{   \frac e {16 \pi^2} \frac v {\sqrt{2} m_\rho^2}}\, {\rm Im}\left[D_L^\dagger \cdot f_{Ld}\cdot Y^D \cdot \tilde{Y}^{D\dagger}\cdot Y^D \cdot f_{Rd}^\dagger \cdot D_R\right]_{11}
\label{nda}
\end{eqnarray}
and similarly for the up quark. 

Using the naive quark model the EDM of the neutron can be estimated in terms of its constituents as,
\begin{equation}
d_n\approx\frac 1 3 (4d_d-d_u)
\label{naiveedm}
\end{equation}
The contribution of the up quark relative to the down quark is further suppressed by the ratio $m_d/m_u\sim 3$ 
so that its contribution is naturally a factor ten smaller than the one of the down quark. With this procedure we get a 
rough estimate,
\begin{equation}
m_\rho\gsim 4\, Y^D\,{\rm TeV}.
\end{equation}
Summarizing, bounds from CP odd observables require a compositeness scale of at least 10 TeV. 
Further, the problem can not be solved by merely aligning the new physics flavor violation with the down sector 
(to avoid the constraints from $\epsilon_K$  and $\epsilon'/\epsilon$), one must also alleviate flavor 
diagonal CP violation to survive the EDM constraints.

\section{Flavor Symmetric Composite Higgs}
\label{MFV}

To make progress within the framework of partial compositeness it seems clear that assumptions 
must be made on the strong sector or on the mixings (see appendix \ref{5dmodel} for the realization in five dimensions). 
In this section, we abandon the mechanism of partial compositeness to suppress
flavor transitions of composite Higgs models and show that if the strong sector has trivial flavor structure,
it is possible to realize the hypothesis of Minimal Flavor Violation \cite{MFV1,MFV}. Regrettably this assumption is incompatible with the dynamical mechanism 
to generate flavor hierarchies starting with structure-less or anarchic 
Yukawas, which is realized in Randall-Sundrum scenarios or theories with a conformal sector \cite{flavorgeneration}. Flavor is, as in the SM, generated  
in the UV, external to the strong sector, and decoupled from the electro-weak scale. The advantage of this approach is however that
the flavor suppression is sufficiently effective for all observables with a compositeness 
scale of $\sim 3$ TeV which is demanded by the $S-$parameter and it applies in general to models 
where the Higgs is a composite state even if the dynamics is not conformal.

The hypothesis of MFV  can be realized within the partial compositeness paradigm 
as follows.\footnote{See \cite{Cacciapaglia:2007fw,duccio}) for related work and \cite{rattazzizaffaroni,perez} for a different construction 
which assumes that the strong sector has flavor dynamics which however is controlled only by two structures, identified with the SM Yukawas.}  
We require that the composite sector respects a flavor symmetry which forbids any flavor violation but still allows the presence 
of degenerate Yukawas. This symmetry should be unbroken in the vacuum. In the most economical realization  we consider a composite sector which possesses an $SU(3)$ flavor symmetry under which  the composite fermions transform as fundamentals.  The composite sector Yukawas and masses in eq. (\ref{2sitecustodial}) must be proportional to the identity. The flavor structure of the SM must be generated by the mixing parameters. General mixings will introduce new flavor structures beyond the ones present in the SM (the CKM matrix and the masses) and this will potentially lead to problems with FCNCs. MFV can now be realized assuming that the left mixings are trivial while the right mixings are  proportional to the SM Yukawas,
\begin{eqnarray}
\lambda_{Lu}\propto Id\,,~~~ \lambda_{Ld} \propto Id\nonumber \\
\lambda_{Ru}\propto y_u\,,~~~ \lambda_{Rd} \propto y_d
\label{eqcompleft}
\end{eqnarray}
We refer to this scenario as left-handed compositeness.
When $y_{u,d}=0$ the theory has a global symmetry $U(3)^3$ which is broken explicitly to $U(1)_B$ by the mixings of right-handed quarks.
This guarantees that as in the SM the only flavor violating structure is given by the CKM matrix. To see this explicitly we can use the singular 
value decomposition (\ref{lambdadecomposition}) for the right mixings. The matrices on the right hand side can be eliminated with a redefinition 
of the elementary right-handed quarks while one of the two matrices on the left can be reabsorbed through a flavor symmetry rotation which leaves the
rest of the lagrangian invariant. The only flavor violating structure is then, as in the SM, the CKM matrix $U_L^\dagger D_L$.
This shows that the theory realizes MFV.

If we assume that the flavor structure is contained in the left mixings we would not obtain a theory which respects MFV\footnote{In order for the flavor structure to arise from the left mixings it is necessary that the elementary left fields mix to more than one fermion species of the strong sector, as is the case
in our scenario.}.  Indeed in this case the diagonal right mixings would break the strong sector + SM flavor symmetry to $SU(3)^2$ which is 
insufficient to take the left mixings to the CKM form. As a consequence additional flavor violation would be generated in this case.
To realize this type of scenario we note that, because of the $U(1)_X$ symmetry, the $SU(3)$ flavor symmetry of the action (\ref{2sitecustodial}) accidentally implies an $SU(3)^2$ symmetry under which composite quarks which couple to up and down SM quarks rotate independently. We may then naturally consider scenarios where this is a symmetry of the composite sector. In this case the MFV hypothesis can be realized in various ways. 
A simple option on which we will focus is indeed that the flavor structure originates from the left mixings and composite right-handed quarks,
\begin{eqnarray}
\lambda_{Lu}\propto y_u\,,~~~ \lambda_{Ld} \propto y_d \nonumber \\
\lambda_{Ru}\propto Id\,,~~~ \lambda_{Rd} \propto Id\,.
\label{eqcompright}
\end{eqnarray}
Note that if the right-handed quarks are part of the composite sector and no elementary partner exists, this choice automatically 
follows from the global symmetry of the strong sector.
Alternatively it is possible to construct  mixed scenarios where the up Yukawa is given by
the left mixings while the down Yukawa is given by the right mixings or vice versa.
In all cases the diagonal mixings leave an $U(3)^3$ global symmetry unbroken.
When this symmetry is explicitly broken by the mixings proportional to the Yukawas the only flavor structure is
the CKM matrix realizing MFV.

In a strongly coupled theory, we expect resonances associated to the global symmetries, 
$SU(3)$ or $SU(3)_U\otimes SU(3)_D$, which are created acting with the respective currents on the vacuum.
This is for example the case in the 5D realization of these models, see \cite{Cacciapaglia:2007fw}. The effective lagrangian
(\ref{2sitecustodial}) should also include these states. Note that since the flavor symmetries are neither weakly gauged nor 
spontaneously broken there are no extra light degrees of freedom but we expect them to have a mass similar to the electro-weak
resonances.

We have shown how MFV can be realized in CHM with one Higgs doublet.
However the flavor protection that we discussed  works in general within the framework 
of partial compositeness and could for example be applied to technicolor or Higgsless  models \cite{higgsless}. 
The electro-weak constraints  are more severe in this case since the scale of compositeness is 
necessarily less than 3 TeV but the flavor problem would be solved.

Another application are CHMs with an extended Higgs sector. In the most compelling composite scenarios 
the Higgs is a Goldstone boson associated to the spontaneous  breaking of a global symmetry of the strong sector, 
$SO(5)/SO(4)$ in the simplest realization \cite{minimalcomposite}. In absence of the microscopic 
theory that realizes the pattern of symmetry breaking,  
there is no a priori reason to prefer a single Higgs model with respect to a multi-Higgs scenario. Moreover 
the fine tuning does not necessarily worsen. This possibility was studied in detail in \cite{cthdm}, 
where various cosets leading to two pseudo-Goldstone  Higgs doublets were considered.

The flavor problem is ubiquitous  in two Higgs doublet models and is even more severe in composite scenarios due
to the non-renormalizable nature of the effective theory which allows more flavor violating operators. 
Interestingly within the MFV hypothesis the flavor problem is solved identically to the single Higgs case because
the flavor symmetry forbids any flavor violation in the composite sector and the SM flavor structure is generated by the mixings.
To see how this works in an example consider the coset \cite{cthdm},
\begin{equation}
\frac {SO(6)}{SU(2)_L\otimes SU(2)_R\otimes U(1)}
\end{equation}
which delivers two Higgs doublets in the $(2,2)$ of custodial. Fermions can be embedded 
in the 6 reps which decomposes as $(2,2)_0+1_2$. As explained in \cite{cthdm} several strong sector operators producing 
Yukawas can be written down. Nevertheless assuming our flavor symmetries the flavor structure is contained in the mixings. 
As a consequence if two of them respect the flavor symmetry, the theory realizes MFV  and
flavor transitions are suppressed to an acceptable value. The same scenarios with left-handed and right-handed compositeness
can be realized in this case.

\subsection{Constraints}

The motivation to realize MFV in partially composite Higgs models is that flavor bounds are automatically respected. 
Actually this requires some qualification in our scenarios. In general even with
MFV, successful suppressions of flavor transitions requires the scale associated to certain
operators to be larger than $\sim5$ TeV (see e.g.~\cite{MFV}), which is larger than the compositeness scale that we will consider. 
Fortunately in our case there are no FCNCs generated at tree level (see appendix \ref{app4fermiop}), 
while flavor violating effects with up and down type quarks are strongly suppressed with respect to the SM~\cite{Cacciapaglia:2007fw}.
As a consequence these models are consistent with all flavor bounds for any reasonable scale of compositeness.

Having solved the flavor and CP problems, the main constraints arise from precision measurements and
compositeness bounds, to which we now turn.

\subsubsection{Modified Couplings}

Contrary to the anarchic scenario where only the mixing of $b_L$
is large, some of the light quarks have sizable mixing due to the flavor symmetry 
which relates their mixing to the ones of third generation quarks.  
This leads to a modification of the couplings of the light generations which similarly to the
one of the $b$ is very constrained.

The hadronic width and the $b$ partial width of the $Z$ are measured with per mille precision at LEP~\cite{pdg}.
Experimentally
\begin{equation}
R_b =\frac {\Gamma(Z\to b \bar{b} )} {\Gamma(Z \to q \bar{q} )} = .21629\pm .00066
\end{equation}
The SM value is 1$\sigma$ below the LEP measurement. From eq. (\ref{deltagl}) it follows that the mixing 
reduces the coupling at least for the simple choices of representations that we are considering. 
As a consequence, $R_b$ constrains $\delta g_{Z\to b_L \bar{b}_L}/g_{Z \to b_L \bar{b}_L}$ to be less than 2.5 per mille, allowing a 2$\sigma$ deviation from the experimental value. As explained in section \ref{choicerep}, in our models there is no shift of the $b$ coupling associated to the up left mixing because of $Z_2$ symmetry. The bound on $R_b$ is then easily satisfied since the shift of the couplings is controlled by $\lambda_{Ld}$ which is naturally small.
In general all our bounds will arise from the up sector where sizable shifts are possible.

For the total hadronic width we have~\cite{pdg},
\begin{equation}
R_h =\frac {\Gamma(Z\to q \bar{q} )} {\Gamma(Z \to \mu \bar{\mu} )}=  20.767\pm .025
\end{equation}
which is again 1$\sigma$ above the SM value. 
At tree level in the SM we have,
\begin{equation}
\Gamma(Z\to \bar{q}q) \propto 2(g_{Lu}^2+g_{Ru}^2) + 3 (g_{Ld}^2+g_{Rd}^2).
\end{equation}
While the mixings in the down sector are naturally small some of the light quark in the up sector will have large mixings in order  to reproduce the top mass and these are strongly constrained by $R_h$. The relative change in the hadronic width due to a common variation of the couplings in the up and down 
sector is,
\begin{eqnarray}
\frac {\delta R_h}{R_h}=\frac {4 g_{Lu} \delta g_{Lu}+4 g_{Ru} \delta g_{Ru}+6 g_{Ld} \delta g_{Ld}+6 g_{Rd} \delta g_{Rd}} {2(g_{Lu}^2+g_{Ru}^2) + 3 (g_{Ld}^2+g_{Rd}^2)}
\nonumber\\
\approx .57\frac {\delta g_{Lu}}{g_{Lu}}+.11 \frac {\delta g_{Ru}}{g_{Ru}}+1.28 \frac {\delta g_{Ld}}{g_{Ld}}+.04 \frac {\delta g_{Rd}}{g_{Rd}}
\end{eqnarray}
Since as before the corrections to the couplings reduce the SM result, 
allowing again for a 2$\sigma$ deviation from the experimental value we find,
\begin{equation}
\frac {\delta g_{Lu}}{g_{Lu}}< .002
\label{deltaglu}
\end{equation}
The allowed variation for up right coupling is instead about 1\%. 
Note that the constraint on an overall shift of the coupling of left-handed down type quarks is even stronger than the one
on $R_b$, while the right-handed coupling of the down quarks is  weakly constrained due to the smallness of its value in the SM. 
For the latter the bound originates from forward-backward asymmetry of the $b$ which 
actually favors a percent increase of the coupling. 

Significant bounds on the left-handed couplings arise also from the unitarity of the CKM matrix.
The constraint is particularly severe for the first row~\cite{pdg},
\begin{equation}
|V_{ud}|^2+|V_{us}|^2+|V_{ub}|^2= .9999 \pm .0012
\end{equation}
where again the error corresponds to a 2$\sigma$ deviation with respect to the central value.

In our model we obtain from eq. (\ref{operatorzbb}),
\begin{equation}
|V_{ud}|^2+|V_{us}|^2+|V_{ub}|^2 \approx 1- 2\, \delta g_{Lu}=1- .7 \, \frac {\delta g_{Lu}}{g_{Lu}}
\end{equation}
which implies a bound similar to eq. (\ref{deltaglu}). 

Let us also mention precision electro-weak constraints, see \cite{silh} for details. Bounds on the $S$ 
parameter require very generally the scale of compositeness $m_\rho$ to be 3 TeV or larger.
For the $T$ parameter, since the right mixings do not break custodial symmetry, the contribution depends on the 
left mixing. For one fermion species this can be estimated as,
\begin{equation}
\alpha_{em} T\sim \frac 3 {16\pi^2} Y_U^4 \sin^4 \varphi_{q_L} \frac {v^2}{m_\rho^2}.
\label{extT}
\end{equation}
which holds as long as the mixing is small. 

\subsubsection{Compositeness Bounds}
\label{secompbound}

Since some of the light quarks are significantly composite, unlike in the anarchic scenario, 
compositeness bounds from flavor preserving 4-Fermi operators can be important in our scenarios. 

The effective lagrangian of 4-Fermi interactions associated to compositeness 
is commonly parametrized as follows \cite{peskinlane},
\begin{equation}
{\cal L}_{4-Fermi}=c_{LL} \,(\bar{q}_L \gamma^\mu q_L)^2+ c_{RR}\, (\bar{q}_R \gamma^\mu q_R)^2+2 c_{LR}
(\bar{q}_L \gamma^\mu q_L)(\bar{q}_R \gamma^\mu q_R)
\label{4fermiexp}
\end{equation} 
where $q= (u,d)$ and color indices are contracted in each quark bilinear. 
Recent studies \cite{LHCcompositeness} constrain 
$c_{LL}<\,0.2\, \rm{TeV}^{-2}$, which will improve in the future with higher luminosity and particularly higher energies\footnote{We use the most conservative experimental bound obtained with 36 $pb^{-1}$ of luminosity.}. Since QCD does not distinguish left and right, 
the same bound applies to $c_{RR}$ if up and down quarks have equal degree of compositeness, while the 
bound on the LR operator is not reported in the experimental studies, though we do not expect it to differ much.  

As shown in the appendix \ref{app4fermiop}, the exchange of composite gauge bosons produces effective 4-Fermi operators whose structure 
is in general different from (\ref{4fermiexp}). For example exchange of gluon resonances generates the flavor diagonal operator,
\begin{equation}
\frac {g_\rho^2}{4\, m_\rho^2}\, \sin^4 \varphi_{q_L}\,\left(\bar{q}_{L\alpha}^i \gamma^\mu  q_{L\beta}^i \bar{q}_{L\beta}^j \gamma_\mu  q_{L\alpha}^j\right)
\end{equation}
where greek indexes are color and latin flavor, and similarly for right-handed compositeness, with the possibility of different mixings for the
up and down right-handed quarks. Nevertheless it is not hard to obtain an estimate of the bounds in our models\footnote{We are grateful to Javi Serra
for discussions about this point.}.   Due to the fact that the bound on the compositeness scale originates from the region of highest dijet invariant mass 
only operators containing valence quarks of the proton contribute significantly. As a consequence, the effect of the operator above has similar effect to (\ref{4fermiexp}) with,
\begin{equation}
c_{LL}\sim\frac {g_\rho^2}{4\, m_\rho^2}\, \sin^4 \varphi_{q_L}
\end{equation}
Adding conservatively similar  contributions due to the exchange of flavor gauge bosons  we find,
\begin{equation}
\sin^2 \varphi_{q_{L,R}}\lsim \frac 2 {g_\rho} \left(\frac{m_\rho}{3 \,\rm{TeV}}\right)
\label{4fermibound}
\end{equation}
respectively for left-handed and right-handed compositeness. 

Finally LR operators with light quarks are much more suppressed, since the product of left and right mixings 
is set by the quark masses. The only exception is an operator, say for right-handed compositeness, $u_R\bar{u}_R t_L\bar{t}_L$, 
whose coefficient can be estimated as $1/m_\rho^2$, which is weakly constrained.
\\

Let us see how the different scenarios comply with the above constraints.

\begin{itemize}

\item{\emph{Left-handed Compositeness}}~~~~   ($\lambda_{Lu}, \lambda_{Ld} \propto Id$)

Using eq. (\ref{deltaglest}) the bound form the hadronic width requires,
\begin{equation}
\frac {Y v}{\sqrt{2}}\sin \varphi_{u_L}\lsim \frac  {m_\rho}{35}
\label{compositenessbound1}
\end{equation}

To reproduce the top mass we then need,
\begin{equation}
\sin \varphi_{t_R}\gsim  35 \frac {m_t}{m_\rho}.
\end{equation}

This rough estimate indicates that these models can only be consistent with the bound from the hadronic width 
if the compositeness scale is somewhat larger than 3 TeV\footnote{In our estimates we use the as a reference value the top mass at 3 TeV (135 GeV in the $\overline{MS}$ scheme).}. For the lowest scale of compositeness, $t_R$ must be part of the composite
sector so that $\sin \varphi_{u_L}\sim \lambda_t /Y$.

The coefficient of 4 Fermi LL operators in eq. (\ref{4fermiexp}) can be estimated as,
\begin{equation}
c_{LL}\sim \frac {.05}{\rm{TeV}^{2}} \frac {g_\rho^2}{Y^4} \left(\frac {m_\rho}{3\, \rm{TeV}}\right)^2
\end{equation}
which is much smaller than the experimental bound and is suppressed at strong coupling. 
This effect will at best be visible at the LHC with high energy and fhigh luminosities.

Concerning corrections to the $T$ parameter the estimate (\ref{extT}) gets multiplied by 3 since 
up and charm have equal degree of compositeness as the top. Nevertheless because the left-handed mixing  
must be small the correction to $T$ will be rather small.

\item{\emph{Right-handed Compositeness}}~~~~   ($\lambda_{Ru}, \lambda_{Rd} \propto Id$)

The right-handed couplings are measured with 1\% precision by the hadronic width of the $Z$ and so scenarios with composite right-handed quarks 
are in general  more easily in agreement with the data. As explained in section \ref{choicerep} in our setup
(which however is quite generic) the corrections to the right-handed couplings are zero at leading order so that 
we do not obtain significant bounds from the hadronic width measurement.

The most relevant constraint arises from the compositeness bound (\ref{4fermibound}).
Taken at face value for $m_\rho\sim 3 $ TeV, this bound already disfavors full compositeness if $g_\rho$ is large, however
$\sin q_R\sim .8$ is still allowed. Given the uncertainties of our estimates there is however room for the right-handed quarks to be fully composite for $g_\rho \lsim 3$. A lower bound on the mixing follows of course from the necessity of generating the top mass,
\begin{equation}
\sin  \varphi_{q_{R}} \gsim \frac {\lambda_t} Y 
\label{lowerboundphir}
\end{equation}
Note that in this limit $t_L$ is fully composite. If the right-handed quarks are not fully composite then $t_L$ is mostly composite and 
we expect from  eq. (\ref{extT}) a sizable contribution to the $T$ parameter. Given the expected positive contribution to $S$
which we allow, this contribution is welcome, as long as it is positive .
The sign of the correction is model dependent being positive when, after electro-weak symmetry breaking, the doublet 
mixes mostly with an electro-weak  singlet, see for example \cite{gillioz}.

We see that this scenario is more weakly constrained than the one with composite left-handed quarks 
even for a significant compositeness of the right-handed quarks and could be easily  compatible with all measurements for 
a compositeness scale of  $3~$TeV.  The general expectation is that the right-handed fermions 
are strongly composite. It is also conceivable, and compatible with compositeness bounds,
that the right-handed quarks are part of the strong sector, see appendix \ref{app4fermiop}. 
As we will see strong compositeness of SM right-handed quarks could give striking signal at the LHC 
where copious production of resonances of the strong sector would be obtained.
Moreover deviations from the SM predictions for dijet mass measurements could give a first evidence  
for this model.

\item{\emph{Mixed Compositeness}}~~~~  ($\lambda_{Ld}, \lambda_{Ru} \propto Id$)

Under the assumption of smaller down quark mixings the main bound comes from the modification
of the up right coupling. This scenario is then similar to the previous case. The alternative mixed
choice $\lambda_{Lu}, \lambda_{Rd} \propto Id$ on the other hand resembles the scenario with left-handed 
compositeness.

\end{itemize}

\subsection{EDMs}

The hypothesis of MFV eliminates large flavor violating effects which for certain 
CP violating observables, most notably $\epsilon_K$, are in conflict with the data for the scale
of compositeness suggested by the hierarchy problem. One still has to worry about flavor diagonal 
phases which may contribute to EDMs and are in general not sufficiently suppressed by the MFV assumption.

Due to the flavor symmetries masses and Yukawa couplings of the composite sector lagrangian in eq. (\ref{2sitecustodial}) are 8
complex parameters (similar arguments hold for the scenario where each fermion couples to a single operator).  
We can independently rephase  the 8 chiral fields. Since two $U(1)$ rotations remain unbroken by  the masses and Yukawas (corresponding to 
baryon number in the up and down sector)  6 phases can be removed and we are left with two diagonal CP violating phases.

It is easy to see that the two new phases remain physical when we introduce the SM fermions. Let us focus on the scenario
$\lambda_{Lu}, \lambda_{Ld}\propto Id$. From eq. (\ref{2sitecustodial}) when $\lambda_{Ru,Rd}=0$ the lagrangian is invariant 
under $U(3)^3$. The right mixings break this symmetry to $U(1)_B$. By a flavor rotation they can be rotated to the standard 
CKM form with 10 physical parameters, in addition to the ones of the composite sector.

A systematic way to count the physical parameters is the following. The total number of parameters in the lagrangian,
divided into real parameters and phases is
\begin{equation}
(10,10)+ (18,18)=(28,28)
\label{parameters}
\end{equation}
where the first set corresponds to the masses, Yukawas and diagonal mixings while the second are the general right mixings. 
These parameters break the following number of symmetries, 
\begin{equation}  
(9,15)+(0,11)-(0,1)=(9,25)
\label{brokensym}
\end{equation}
The first factor corresponds to $SU(3)^3$, the second to the 11 $U(1)$  
associated to rotations of the chiral fermions (composite and elementary) 
and the last is the unbroken baryon number. The difference between eq. (\ref{parameters}) and
eq. (\ref{brokensym}) is the number of physical parameters,
\begin{equation}
(19,3)
\end{equation}
The 19 real parameters correspond to 4 masses and 4 mixings angles
of the heavy fermions (induced by the composite sector Yukawas), 6 SM fermion masses, 3 angles of the CKM matrix and the left universal mixings $\lambda_{Lu}, \lambda_{Ld}$. 
The 3 phases are the CKM phase and the two new phases of the strong sector.

The extra phases mediate new CP violating effects \cite{straub}. We can always choose a basis where 
the couplings $Y$ are real. With a flavor rotation we can take,
\begin{eqnarray}
&\lambda_{Ru} = \hat{\lambda}_u V\nonumber \\
&\lambda_{Rd} = \hat{\lambda}_d
\end{eqnarray}
where $ \hat{\lambda}_{Ru,Rd}$ are diagonal and real and $V$ is a unitary matrix. 
By $U(1)$ rotations on $u_R$ and $U_L$ the matrix $V$ can be taken to the standard CKM form with one physical phase. 
The $Y$'s and $\lambda_{Lu,Ld}$ are complex at this stage. $\lambda_{Lu}$ can be made real by rephasing $q_L$, 
$\lambda_{Ld}$ by rotating $Q_{Rd}$ while $Y_u$ rotating $Q_{Ru}$ and $Y_d$ by rotating $D$. 
The latter requires a compensating rotation on $d_R$ to keep $\hat{\lambda}_d$ real. In this basis $Y$ are real and 
$\tilde{Y}$ are complex. Since the leading order contribution to dipoles (\ref{nda}) depends on $\tilde{Y}$ 
it follows that  an unsuppressed contribution to EDMs from loops of composite fermions is generated.

The IR contribution to EDMs could be reduced assuming $\tilde{Y}\ll Y$.  
Even so, in complete models we expect an unsuppressed UV contribution that would cause tension
with EDMs bounds. For example the operator in the composite sector,
\begin{equation}
{\cal L}_{UV}=\frac {g_{\rho}^2}{(4\pi)^2}\frac {g_\rho} {m_\rho^2}\, {\rm Tr}[\bar{L}_{U}  {\cal H}] \sigma^{\mu\nu}U\,e  \tilde{F}_{\mu\nu}\,.
\end{equation}
upon mixing with the elementary fields generates an up quark EDM. With the NDA estimate above this contribution is of similar size 
as unsuppressed fermionic loops. Problematic UV and IR contributions to EDMs can be eliminated assuming that the strong sector is CP symmetric. 
In this case there is a single physical CP violating phase (the CKM phase) induced by the elementary-composite mixings.
As a consequence, if the strong sector is CP invariant the UV and IR contributions to dipoles do not generate EDMs to leading order.

\subsection{LHC phenomenology}

In this section we outline the basic features of the phenomenology of MFV scenarios presented above.
We adopt the simplifying assumption that all the couplings of the strong sector are equal so that for example $Y= g_\rho$. 
Many conclusions could rapidly change even for small hierarchies between different couplings and moreover
a thorough analysis at the LHC should be performed to determine the discovery reach. 
With this in mind we wish to stress that experimental prospects to directly produce the new states 
associated to the strong sector are generically more promising that in the anarchic scenario, 
due to the enhanced coupling of the SM fermions to strong sector resonances.

Experimentally the most notable difference compared to the scenario with anarchic strong sector Yukawas 
is that the heavy composite states have large couplings to at least some of the light quarks. 
This is true in particular in the second scenario where right-handed fermions are substantially or even completely composite.
This changes  the phenomenology enormously allowing in certain cases to easily produce the new states at the LHC, 
possibly changing the experimental strategies due to different decay channels.

Production of composite resonances within partial compositeness scenarios has been studied
in \cite{gluonresonance,agashe,singleproduction,doubleproduction} with emphasis on minimal Randall-Sundrum models. 
Our models do not require any type of conformal dynamics or extra-dimensions. Given that our setup-up 
cannot aim to the dynamical generation of  flavor structure, this additional structure is superfluous. As a consequence
the parameter space that we consider is larger than in the studies above.

\subsection{Spin 1}

Let us begin our discussion considering spin-1 resonances of SM gauge bosons. 
Following \cite{2sitescontino} the trilinear coupling of SM fermions to strong sector resonances is given by
\footnote{For the left fermions which couple to two different composite fermions we will consider only the larger contribution
associated to $\lambda_{Lu}$.},
\begin{equation}
g_{\rho\bar{\psi}_\varphi \psi_\varphi}= g (\sin^2 \varphi \cot \theta - \cos^2 \varphi \tan \theta) 
\label{spin1coupling}
\end{equation}
where $g$ is the SM coupling, $\varphi$ is the mixing angle of the relevant fermionic chirality,
while $\theta$ is the mixing angle of SM gauge bosons and spin 1 resonances. This interaction originates from the rotation 
in eq. (\ref{2sitecustodial}) of the elementary/composite states to the mass basis, the first term corresponding 
to mixings of the fermions while the second to mixing of the vectors (analog of photon-$\rho$ mixing in QCD).

In Randall-Sundrum models without boundary kinetic terms one finds
\begin{equation}
\tan \theta = \frac {g_{el}}{g_\rho} \sim \frac 1 5\,,~~~~~~~~~~~~~~~~~~~~~~ g_{el} \sim g
\end{equation}
which is determined by the Planck-weak hierarchy. This choice, assumed in Refs. \cite{gluonresonance,agashe}, 
corresponds to $g_\rho\sim 5$ for the resonances associated to $SU(3)_c$ and $g_\rho\sim 3$ for $SU(2)_L$. 
We assume here that the couplings of the strong and elementary sectors are independent. 
For simplicity, and in agreement with the logic that the strong sector is characterized by interactions with strength $g_\rho$,
we also assume equal coupling for all the resonances which implies a smaller mixing angle for electro-weak 
resonances. 

The assumption that the strong sector couplings are independent could have an important experimental impact 
at the LHC. In particular reducing the coupling of the composite sector favors the production of the new states. 
This slightly counter-intuitive conclusion follows from the fact that a large coupling requires a small mixing which 
enters quadratically in the coupling to strong sector resonances. 

In the anarchic scenario the first term in eq. (\ref{spin1coupling}) can be neglected, and the coupling 
is  determined by the mixing of SM gauge fields with vector resonances. 
This does not necessarily apply in our scenario with composite light quarks, the first contribution becoming more important for,
\begin{equation}
\varphi \gsim  \theta
\end{equation}

Let us now discuss the consequences for left-handed and right-handed compositeness.

\subsubsection{Left-handed Compositeness}

For $m_\rho\sim 3$ TeV,
\begin{equation}
\sin \varphi_{u_L} \sim  \frac {\lambda_t} Y.
\label{boundul1}
\end{equation}
Substituting into eq. (\ref{spin1coupling}) one finds,
\begin{equation}
g_\rho\left(\frac {\lambda_t^2} {Y^2}-\frac {g_{el}^2}{g_\rho^2}\right)
\end{equation}
For $g_\rho\sim Y$ we see that the two terms are comparable for gluon resonances ($g_{el}\sim 1$)
and moreover they have opposite signs. As a consequence we do not expect production cross sections larger than the ones in the 
anarchic scenario.  For electro-weak resonances however, since $g_{el}\sim 0.5$, the first term could dominate and  larger cross-sections
could be obtained. Recall that in this scenario for low compositeness scale the right-handed top is 
fully composite while the (universal) left mixing scales as $1/Y$. As a consequence, unless $Y$ is close to 1, 
the gluon resonances will predominantly decay into right-handed tops as in the anarchic scenario. 
Note that the width of the gluon resonances (for equal coupling) is not much larger even though their production rate is. 

\subsubsection{Right-handed Compositeness}

The situation is much more promising in this scenario. We focus on gluon resonances, which are heavy color octets. 
For the smallest mixing allowed to reproduce the top mass, $\sin\varphi_{q_R}\sim \lambda_t/Y$, 
the analysis reduces to the one above, with minor variations if the compositeness 
of up and down quarks is different. For larger mixings, which we generically expect, production of resonances due to fermion 
mixing certainly dominates and very large cross-sections can be obtained. 

\begin{figure}[t]
\begin{center}
\includegraphics[scale=0.42]{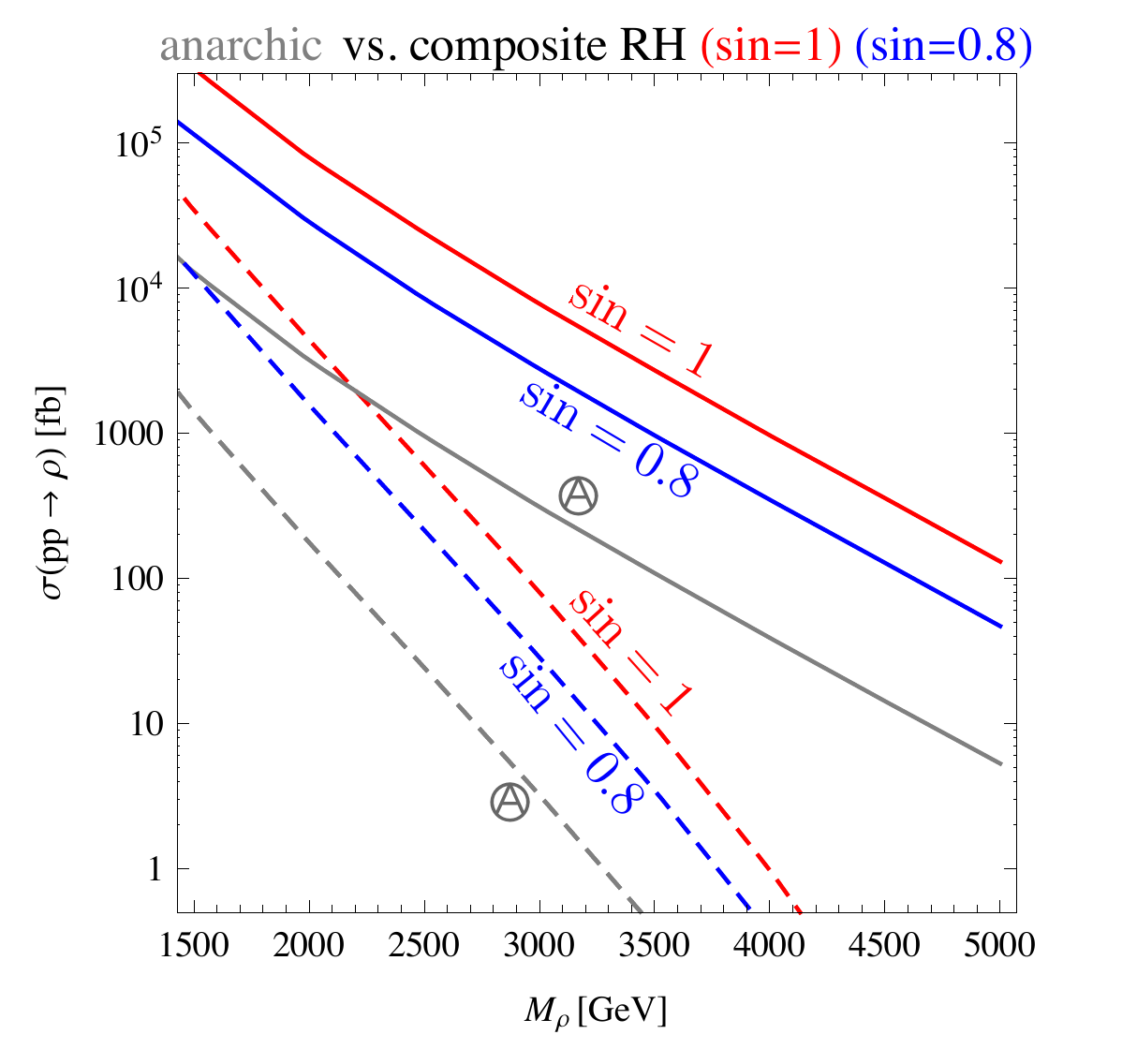}\hspace{.2cm}
\includegraphics[scale=0.9]{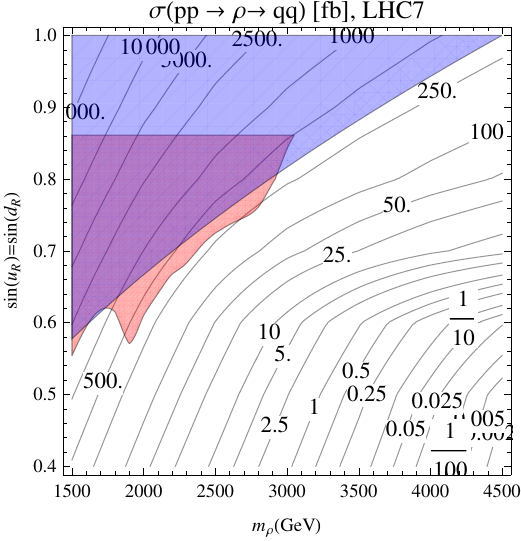}
\caption{a) Production cross-section for $p p \to \rho$ of the gluon resonance at LHC7 (dashed) and LHC14 (solid) for $g_\rho=3$ obtained with Madgraph~\cite{madgraph}.  The gray line corresponds to the anarchic scenario and the red line to the scenario with fully composite right handed fermions. 
b) Cross-section for $p p \to \rho\to q_i q_i$ with $i=u,d,c,s,b$ as a function of the $\rho$-mass and the right-handed mixing angles. We show the constraint from compositeness (blue) and the region excluded by the dijet resonance search (red). See text for details.}
\label{fig:crossx}
\end{center}
\end{figure}

In Fig.~\ref{fig:crossx}a we report the production cross-section of gluon resonances at LHC 
with 7 and 14 TeV for composite right-handed quarks. The fully composite case shows the largest cross-section that can be 
obtained in our scenario. Signals for our model should be seen early, especially at LHC14. 
Note that full compositeness is disfavored, especially in the low mass region, by compositeness bounds 
and direct searches (see below). Also in the very low mass region direct searches for $t\bar t$ resonances already  
place a mild bound on the model \cite{atlasdirectbound} for $m_\rho< 2$ TeV. 
In figure~\ref{fig:crossx}b we show the currently allowed parameter space in the plane of the quark mixing and mass of the resonances. 
We use  the ATLAS dijet resonance search~\cite{LHCcompositeness}  and its recent update \cite{dijet163invpb} (using 161 pb$^{-1}$ of data). 
As can be seen in figure~\ref{fig:crossx}b, the dijet resonance constraints are at present similar or even slightly stronger
than the compositeness bounds for masses up to $\sim 3$ TeV. 

For the direct search of resonances we extract the constraints on our model using the following procedure.\footnote{We thank Georgios Choudalakis for discussions.}
First, we simulate parton-level events in our scenario with MadGraph/MadEvent~\cite{madgraph} using the CTEQ6L1 PDFs~\cite{cteq}. We replicate the cuts of~\cite{dijet163invpb} on $p_T$, $|\eta|$, and the selection requirement for the two leading jets  ($|\Delta\eta| < 1.3$). The latter reduces the number of events by about a factor of 2. We find acceptances in agreement with the ones mentioned in the ATLAS analysis~\cite{LHCcompositeness}. We then require the invariant mass of the dijets to be in a window of $\pm$20 \%  around the resonance mass. This range is large enough for the widths in the region where the dijet resonance search is competitive. 
We combine the efficiency of the cut around the bump with the previous ones to a total acceptance.  Finally we compare 
$\sigma(pp \to \rho\to \sum_i q_i q_i)\times$acceptance ($i=u,d,s,c,b$) to the limits reported in~\cite{dijet163invpb}. We do not include collimated tops to be conservative since it is hard to estimate their efficiency.
At very large values of the mixing the  bound is not effective anymore since the resonances become too broad and there are currently no 
experimental limits. We expect the constraint on the cross-section in the region of interest to soon improve, and scale 
approximately as $\sim (\mathcal{L}_{new}/\mathcal{L}_{old})^\frac12$.

Even if the right-handed fermions are not fully composite cross-sections larger than in the anarchic scenario are generically
obtained. At LHC 14 where the resonances will be mostly produced on-shell a rough estimate of the total cross-section can be obtained
rescaling the production cross-section of the anarchic scenario. For a 3 TeV gluon resonance we have approximately,
\begin{equation}
\sigma_{MFV}
\sim \, \left[(g_{\rho\bar{\psi}_{uL} \psi_{uL}}^{MFV})^2+(g_{\rho\bar{\psi}_{uR} \psi_{uR}}^{MFV})^2+.5 (g_{\rho\bar{\psi}_{dL} \psi_{dL}}^{MFV})^2+ .5(g_{\rho\bar{\psi}_{dR} \psi_{dR}}^{MFV})^2\right]\, \rm{pb}
\end{equation}
At LHC7 the resonances are mostly off-shell and the production cross-sections will be smaller than the total cross-section, see Fig. \ref{fig:crossx}. In table \ref{tab:xsec14} we collect the production cross-sections of 3 TeV gluon resonances at LHC14 for various choices 
of couplings and mixings obtained with Madgraph/Madevent~\cite{mad graph} using the CTEQ6L1 PDFs~\cite{cteq}.
\renewcommand{\arraystretch}{1.}
\begin{table}[t]
\begin{center}
\begin{tabular}{c|c|c|c|c|c|c|c}
 $g_{\rho}$ & $\sin \varphi_{u_R}$ & $\sin \varphi_{d_R}$ & $\sigma (\rm{pb})$ & $\Gamma$(GeV) &  Br($u\bar{u}$)& Br($t_L\bar{t}_L$) & Br($t_R\bar{t}_R$)\\ \hline \hline
3 & \CircledA & \CircledA & $0.31$ & $190$ & $0.03$ & $0.004$ & $0.84$ \\  \hline
5 & \CircledA & \CircledA & $0.1$ & $480$ & $0.004$ & $0.0005$ & $0.98$ \\  \hline
3 & 0.4 & 0.4 & $0.17$ & $50$ & $0.07$ & $0.35$ & $0.005$ \\  \hline
5 & 0.4 & 0.4 & $0.38$ & $60$ & $0.14$ & $0.1$ & $0.12$ \\  \hline
3 & 0.8 & 0.8 & $2.8 $ & $340$ & $0.17$ & $0.0003$ & $0.16$ \\  \hline
5 & 0.5 & 0.5 & $1.1$ & $140$ & $0.17$ & $0.01$ & $0.16$ \\  \hline
3 & 1 & 0.25 & $5.2$ & $490$ & $0.33$ & $0.001$ & $0.32$ \\  \hline
5 & 1 & 0.25 & $15$ & $1400$ & $0.33$ & $0.0002$ & $0.33$ \\  
\end{tabular}
\end{center}\caption {Production cross-section of 3 TeV spin-1 color octet resonance at LHC14 for the anarchic (\CircledA) and MFV 
scenarios for benchmark values of couplings and mixings.}
\label{tab:xsec14}
\end{table}
\renewcommand{\arraystretch}{1.}

In the anarchic scenario the resonances are mostly coupled to the third generation. In particular if the top right
is part of the strong sector the decay into tops is dominant. As we already mentioned this is not necessarily true in our scenario. 
The width of gluon resonances is given by,
\begin{equation}
\Gamma \simeq \frac {m_\rho}{48 \pi} \sum_i g_{\rho\bar{\psi}_\varphi^i \psi_\varphi^i}^2
\end{equation}
For moderate mixing $t_L$ must be significantly composite in order to  reproduce the top mass 
so that the main decay will be again into third generation states.
However in the regime of large mixings the decay into light quarks will dominate. This happens for 
\begin{equation}
\sin^2 \varphi_{u_R} \gsim \frac {\lambda_t} {Y} 
\end{equation}
which is also the region of parameters where large cross-sections are generated.
As we can see in table \ref{tab:xsec14}, in this case the resonances tend to be broader than in the anarchic scenario due to the multiplicity 
of the decay channels interacting strongly. When right-handed quarks are completely composite this feature forbids very large couplings $g_\rho$. 
One finds that for   fully composite right-handed  quarks the width of the resonances is comparable to the mass for $g_\rho\sim 5$, 
so we take this as the maximal value.
We note that while a large violation of flavor universality would be a hint to the anarchic generation 
of flavor the opposite will be true here: universality would directly indicate the presence of flavor symmetries
in the strong sector. 

Let us now briefly comment on the other vector resonances. For electro-weak resonances  $g_{el}\sim 1/2$ so 
the first term in eq. (\ref{spin1coupling}) will certainly dominate (the same was already true for the case of left-handed compositeness). 
In all cases we obtain cross-sections  which are at least ten times larger than in the anarchic case. 
Note that the coupling will be similar to colored resonances since we are assuming a common coupling $g_\rho$ for
the strong sector. The  decays are more complicated in this sector and we leave a detailed study to future work. Finally 
we also  expect in our theory vector resonances associated to the flavor symmetries of the model. These will be produced
with cross-sections similar to the ones above and their detection would provide a distinctive feature of this model.

\subsection{Spin $1/2$}

\begin{figure}[t]
\includegraphics[scale=0.9]{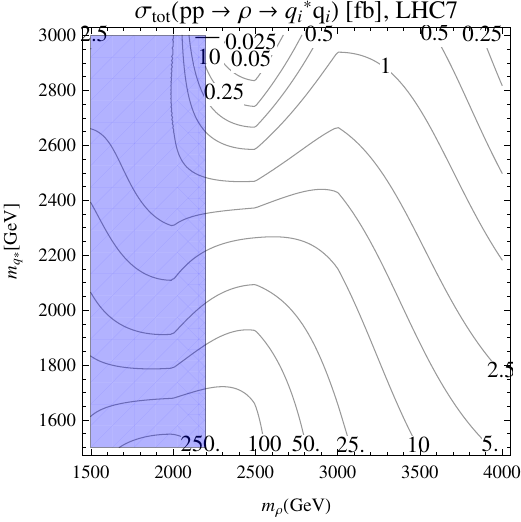}\hspace{.5cm}
\includegraphics[scale=0.9]{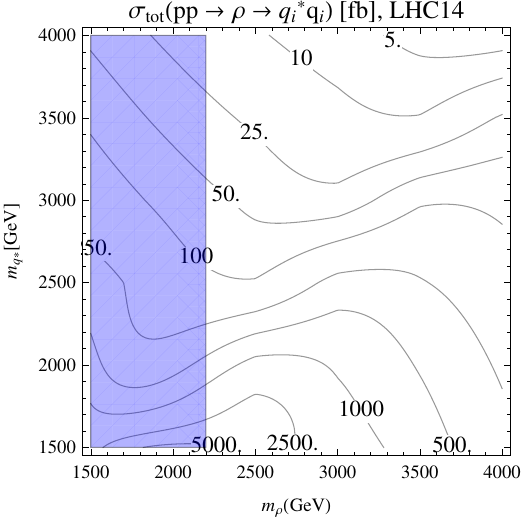}
\caption{Total cross-section of associated production of partners of the first two generations at LHC7 and LHC14 obtained with Madgraph~\cite{mad graph} for $g_\rho =3$, $\sin \varphi_{u_R} = 0.7 $, $\sin \varphi_{d_R} = 1/6 $. The  blue region shows the parameter space excluded by compositeness searches~\cite{LHCcompositeness}. The dijet resonance search does not lead to a constraint for this choice of parameters. }
\label{fig:fermioncrossx}
\end{figure}
\begin{wrapfigure}{r}{0.3\textwidth}
\vspace{-40pt}
\begin{center}
\includegraphics[scale=.45]{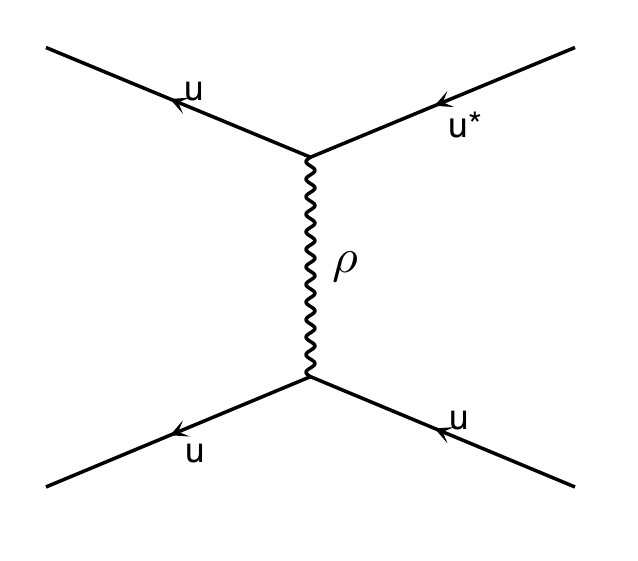}
\end{center}
\vspace{-40pt}
\caption{Dominant contribution to $\sigma(pp\to \rho \to \chi q)$.}
\label{fig:tqqstar}
 \vspace{-0pt}
\end{wrapfigure}

Fermionic resonances can be singly ~\cite{singleproduction} or doubly produced \cite{doubleproduction}. 
In the anarchic scenario the latter process proceeds dominantly through QCD interactions and this contribution will be identical in our case.
The heavy fermions can also be produced through the exchange of heavy spin-1 resonances,
and this process can be relevant in our scenarios. For an off-shell intermediate state, the amplitude scales as,
\begin{equation}
A[q\bar{q}\to \chi\bar{\chi}]\propto g_\rho^2 \left(\sin^2 \varphi-\frac {g_{el}^2}{g_\rho^2}\right)\,.
\end{equation} 
Note, that this formula and the one below are only valid for not extremely large mixings.
In the scenario with right-handed compositeness, the contribution of gluon resonances 
can be easily an order of magnitude larger than QCD production.
As an example at LHC14 we simulated the process $pp\to \chi_3 \chi_3$ with MG/ME 4 for $m_\rho=3$ TeV  
and $m_\chi=2$ TeV and the same mixings as in figure \ref{fig:fermioncrossx}. We find an enhancement by a factor 
of about 4 compared to QCD production. 

Concerning single production, the channel which has been most studied is the scattering of $W_L$ and a top producing a quark
resonance. In our scenario with right-handed compositeness the same process is possible through the scattering of light quarks 
with naturally much larger cross sections. Moreover single production is also possible with the exchange of gluon 
resonances, $pp \to \rho^* \to q \chi$.
\begin{equation}
A[q\bar{q} \to  q \chi]\propto  g_\rho^2 \,\sin \varphi\,  \left(\sin^2 \varphi-\frac {g_{el}^2}{g_\rho^2}\right)
\end{equation}
Compared to  double production this process has a coupling suppressed by the mixing but it is of course more
favorable energetically. The dominant contribution is through the t-channel spin-1 exchange shown in figure~\ref{fig:tqqstar} 
which benefits from  access to the  up quark pdf in the two protons. 
In figure \ref{fig:fermioncrossx} we show the cross-section at LHC14 for single production mediated by a color octet spin 1 resonance. 

In summary, our preliminary study shows significantly enhanced cross-sections compared to the standard partial compositeness scenario which bodes well for the discovery potential of our model. A dedicated study can be found in \cite{Redi:2013eaa}.

\section{CP Invariant Composite Higgs}
\label{CPprotection}

In this section we propose a different mechanism to relieve flavor bounds in CHM, based on the idea that
the strong sector respects CP. As reviewed in section \ref{cpreview}, one intriguing feature
is that practically all phenomenological tensions with experiment originate from CP violating observables.
In general, even though deeply connected within the SM, the origin of CP and flavor breaking are
logically independent. This motivates to look for models where the flavor structure and  CP 
violation in the SM are generated independently, as the mechanism of partial compositeness 
is sufficient to suppress flavor violations of CP even observables within experimental bounds. 
The purpose of this second part of the paper will be to show a realization of this idea. Under suitable assumptions, 
this allows to generate the flavor structure  by RGE while suppressing unwanted contribution to CP violating 
observables.

To separate the generation of flavor structure and CP violation we will make the minimal hypothesis that the strong sector 
is CP symmetric, as QCD is. In the language of the effective lagrangian (\ref{2sitecustodial}) this means 
that the Yukawa couplings of the composite sector can be chosen to be real matrices. 

In order to reproduce the CKM phase of the SM, the mixings must violate CP. 
This implies that the mixing matrices $\lambda_{L,R}$ are complex and cannot be 
made real by a unitary redefinition of the elementary fields. There are many ways to  realize this. 
If at low energy $\lambda_{Lu}$, $\lambda_{Ld}$, $\lambda_{Ru}$,  $\lambda_{Rd}$ are all complex matrices with large phases, 
the contributions to CP violating processes are similar to the ones of the anarchic scenario studied in 
the literature.  To see this, one can use the singular value decomposition for the mixings (\ref{lambdadecomposition}).
The left matrices can be eliminated with a redefinition of the elementary fermion fields while the right matrices can 
be rotated into the Yukawas of the composite sector which become anarchic complex matrices.
This effectively reduces to the situation normally considered in the literature with anarchic complex Yukawas and diagonal mixings.

Interestingly, if flavor is generated dynamically, there is a natural suppression of CP violation induced from the elementary sector.
Consider first the case where each SM quark chirality couples to a single operator.
In the UV the mixings are,
\begin{equation}
\lambda_{ij}(\Lambda_{UV}) \bar{q}^i O^j
\end{equation}
where $\lambda_{ij}(\Lambda_{UV})$ is assumed to be an anarchic matrix with order one entries.
Considering for simplicity the case where the coupling is irrelevant at  low energy we obtain,
\begin{equation}
\lambda_{ij}(\mu_{IR})=\lambda_{ij}(\Lambda_{UV})\left( \frac {\mu_{IR}}{\Lambda_{UV}}\right)^{\gamma_j}
\end{equation}
where $\gamma_i=[O_i]-5/2$. The hierarchies of the second term can naturally generate the hierarchies of quark masses 
and of the CKM matrix. However, the same hierarchies also suppress  CP violation in the SM.
If the eigenvalues of the UV mixings are order one it is possible to show that the CKM phase is highly suppressed.
This can be understood intuitively because when the eigenvalues of $\lambda_{ij}(\Lambda_{UV})$ are equal all complex parameters 
are unphysical and can be reabsorbed through a redefinition of the elementary fields.

As a consequence if each SM fermion couples to a single operator, the generation of the flavor hierarchies 
by RGE is not compatible with obtaining an order one phase of the CKM matrix.

This conclusion changes completely if  SM fermions are coupled to two or more operators. This the case in our scenario
for the left-handed quarks. More generally, this could be realized if some of the SM fermions 
couple to more than one operator. To avoid the complication mentioned in section \ref{cpreview}
we will assume that the coupling of the left-handed elementary fields to be aligned in flavor space.
This can be enforced by a simple UV symmetry of the theory, see e.g.~\cite{Csaki:2008eh}. In this case, 
even though the mixings are diagonal, the relative phase between the up and down sector is physical and is not washed out 
by the running. Note however that CP violation induced from the mixing of right-handed fermions, if present in the UV, 
is still suppressed at low energies as explained above. Using a redefinition of the elementary 
fields we can assume that the phases are in the mixings of the left-handed up sector. 
It is easy to see that if at least one of the phases is large, an order one phase will be induced  in the CKM matrix. 
The Yukawas are,
\begin{eqnarray}
y_u&=& e^{i\vec{\alpha}}. \lambda_{Lu}.Y^U. \lambda_{Ru} \\
y_d&=& \lambda_{Ld}.Y^D. \lambda_{Rd}
\end{eqnarray}
The left diagonalization matrices are,
\begin{eqnarray}
U_L &=& e^{i \vec{\alpha}}.V_{Lu}\nonumber \\
D_L &= &V_{Ld}
\label{rotationCKM}
\end{eqnarray}
where $V_{Lu}$, $V_{Ld}$ are real orthogonal matrices and $e^{i\vec{\alpha}}=\rm{diag}[e^{i \alpha_1},e^{i \alpha_2},e^{i \alpha_3}]$. 
The CKM matrix is then,
\begin{equation}
V^{CKM}= V_{Lu}^T. e^{-i\vec{\alpha}}. V_{Ld}
\label{CKMmath}
\end{equation}
In order to determine the phase in the CKM matrix we consider $J$ 
\begin{equation}
J=V^{CKM}_{12}\,V^{CKM}_{23}\,(V^{CKM}_{13})^*\,(V^{CKM}_{22})^*
\end{equation}
whose imaginary part is the reduced Jarlskog invariant. $J$ has an 
unsuppressed imaginary part if at least one of the phases in  $e^{i\vec{\alpha}}$ is large. 
The ${\cal O}(1)$ phase of the CKM matrix is then naturally generated
and we easily reproduce the SM value $\rm{Im}[J] \sim 3\times 10^{-5}$. We will show this in
numerical examples.

The matrix in eq. (\ref{CKMmath}), is a unitary matrix not in the standard basis used for the CKM matrix in the SM where
real and imaginary parts of 4-Fermi operators correspond to CP preserving and CP violating effects.
In order to properly identify  contributions to CP violation from new physics it is convenient to rotate into the standard SM basis. 
As usual such basis is reached performing $U(1)$ re-phasing on the left-handed fields compensated by opposite rotations on the right fields 
to leave the masses real. The transformation is given explicitly in  appendix \ref{CKMredefinition}.

We now show that in this setup CP violating effects can be suppressed.
For EDMs which are flavor diagonal observables, the contribution of new physics is zero 
at leading order for any choice of the phases. Indeed the expression (\ref{nda})  is manifestly real
and the field redefinition necessary to reach the SM basis does not change this feature since the rotation of the left
chirality of the quark is compensated by the right one\footnote{One might wonder if this might also help with the strong CP problem.
Unfortunately one can check that the phase of the determinant of the fermion mass matrix is complex 
so that, even assuming CP to be a spontaneously broken symmetry of the theory the effective $\theta-$angle is 
different from zero. In other words the mechanism of  Nelson-Barr \cite{nelson}  will not work in our case.}.
This solves the problem of EDMs in CHM.

\begin{figure}[t]
\begin{center}
\includegraphics[scale=1]{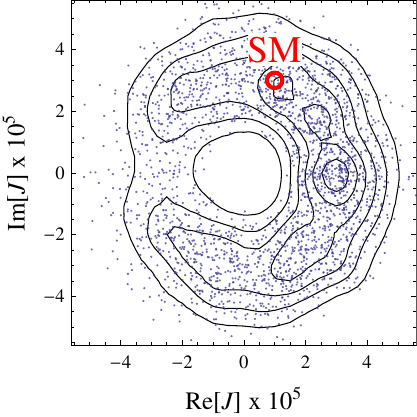}\hspace{.7cm}
\includegraphics[scale=1.08]{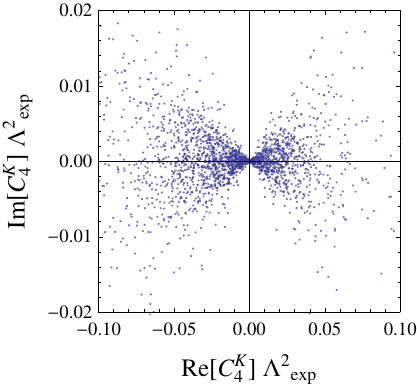}
\caption{On the left we plot real and imaginary parts of the Jarlskog invariant corresponding to different points of the sample of example $g_\rho=3$. The contour lines show the distribution of points.
On the right, we show real and imaginary parts of the Wilson coefficient $C_4^K$ normalized with the respective experimental bound.}
\label{fig:j1}
\end{center}
\end{figure}

The situation is not as simple for flavor violating effects because the phases of different flavors required to reach  
the physical basis of the CKM matrix are unrelated in general. As a consequence, even though the 4-Fermi 
operators obtained in the basis defined by eq. (\ref{CKMmath}) are real, they become complex after the rotation 
to the SM basis, i.e. they contribute to CP violating observables. 
Nevertheless there can be attractive choices. Similarly to the realization of MFV in section \ref{MFV}, 
we can assume that only certain couplings violate CP. 
Phenomenologically it is sufficient to suppress CP violating contribution 
for the light generations. This requires equal phases for the down and strange quarks. 
In the appendix \ref{CKMredefinition} we show that this happens for,
\begin{equation}
{\rm arg}[V^{CKM}_{11}]-{\rm arg}[V^{CKM}_{12}]\ll 1
\end{equation}
in eq. (\ref{CKMmath}).
Remarkably this is automatically realized if CP violation is induced  through the mixing the third generation,
i.e. if CP violation originates from the elementary top and bottom sector mirroring the hierarchy of the mixings,
\begin{equation}
 \vec{\alpha} \sim (0,0,1)\,.
\end{equation}
This can be seen from eq.~(\ref{CKMmath}) as, taking into account the hierarchies of the rotation matrices, both elements are 
approximately real. For the same reason contributions to $\epsilon_K'/\epsilon_K$ are also very suppressed.
Since the phases can be equivalently placed into the down sector, the same conclusion holds for the $D-$system.
On the other hand we obtain no suppression of CP violation for operators involving the third generation. 
In particular the contributions to the effective operators which generate $B_{s,d}$ mixing have an unsuppressed 
imaginary part contributing to CP violation. In these systems the protection from partial compositeness 
is sufficient to satisfy current flavor constraints but improved sensitivity in these channels might be able to detect it.

In this scenario flavor bounds arise from the real parts of $\Delta F=2$ operators and CP odd
observables containing the third generation quarks. 
One finds that $B_d$ and $K$ give a similar bound $\sim$TeV on the compositeness scale.
There is no tension for a compositeness scale $\sim 3~$TeV. As explained in section~\ref{choicerep} there is no sizable correction 
to the coupling of $b_L$ at least as long as $\lambda_{Ld}\ll \lambda_{Lu}$ which we always assume.
In this scenario the most severe bound is  the one from  $b\to s \gamma$ in eq. (\ref{bsgammabound}), 
Note that for this process there are contributions to the real 
and imaginary parts of the operator. This effect is at present beyond experimental reach but for large coupling $g_\rho$ this could put some pressure on the model assuming improved experimental precision.

\subsection{Examples}

To see how effective the mechanism proposed above is, we have generated several samples 
of anarchic composite sector Yukawas that (approximately) reproduce SM masses and mixings (see also \cite{2siteskaustubh}
for a similar study in the anarchic scenario). The mixings are taken to be diagonal with a phase in the third component of $\lambda_{Lu}$.
We do not impose a constraint on the SM CP phase in order to show that an order one phase is naturally generated.

In the first example we choose the following parameters,
\renewcommand{\arraystretch}{1.1}
\begin{center}
\begin{tabular}{|c|c|c|c|c|c|c|c|c}
\hline
$m_\rho$~(GeV) & $g_{\rho}$ & $Y^U$ & $Y^D$ & $\sin \varphi_{t_L}$ & $\sin \varphi_{t_R}$ & $\sin \varphi_{b_L}$ & $\sin \varphi_{b_R}$ \\ 
\hline 
$3000$ & $3$ & $3.1$ & $3.2$ & $0.5$ & $0.75$ & $0.125$ & $0.035$\\ \hline
\end{tabular}
\end{center}
\renewcommand{\arraystretch}{1.}
with the remaining mixing determined in order to reproduce quark masses and CKM angles.
For simplicity we have chosen $Y^U\sim Y^D\sim g_\rho$, where $Y^{U,D}$ are the average eigenvalue of the composite Yukawa matrix.

As shown in figure \ref{fig:j1} there is no suppression of the imaginary part of the Jarlskog invariant.
The average value is $2\times 10^{-5}$, compatible with the SM result.
For the sample of anarchic composite sector Yukawas generated we computed the Wilson coefficient of the
4-Fermi operators for the $K$, $D$ and $B$ systems due to the exchange of gluon resonances. Following Ref. \cite{2sitescontino}
we have assumed that the latter couple identically as gluons but very similar results would be obtained in general. 
The average size of the Wilson coefficients of the effective 4-Fermi operators is reported in  table \ref{table:wilson} together with
the experimental limit (we use the results of \cite{utfit} evaluated at 3 TeV \cite{weiler}). Bounds are easily evaded. In particular the imaginary part of $C_4^K$ is strongly suppressed relative to the real part, see Fig. \ref{fig:j1}.
Note that the imaginary parts of $C_4^{B_s}$ and $C_4^{B_d}$ are not suppressed (with size given by eq. (\ref{LRoperator})) 
but still much smaller than the experimental bounds for this choice of parameters.
\begin{table}[h!]
\begin{center}
\begin{tabular}{c||c|c}
 (in GeV$^{-2}$)  & CP\CircledA\,& EXP \\  \hline
$\rm{Re}[C_4^K]$ 			& $4 \cdot 10^{-16}$ & $7\cdot 10^{-15}$ \\
$\rm{Im}[C_4^K]$ 		& $4\cdot 10^{-19}$ & $4\cdot 10^{-17}$\\ \hline
$\rm{Re}[C_4^D]$ 			& $2\cdot 10^{-15}$ & $8\cdot 10^{-14}$ \\
$\rm{Im}[C_4^D]$ 		& $4\cdot 10^{-18}$ & $8\cdot 10^{-14}$\\ \hline
$\rm{Re}[C_4^{B_d}]$ 			& $5\cdot 10^{-15}$ & $3\cdot 10^{-13}$ \\
$\rm{Im}[C_4^{B_d}]$ 		& $5\cdot 10^{-15}$ & $3\cdot 10^{-13}$\\ \hline
$\rm{Re}[C_4^{B_s}]$ 			& $8\cdot 10^{-14}$ & $2\cdot 10^{-11}$ \\
$\rm{Im}[C_4^{B_s}]$ 		& $8\cdot 10^{-14}$ & $2\cdot 10^{-11}$ 
\end{tabular}
\end{center}
\caption{Average values of Wilson coefficients of effective flavor changing operators in example 1 and 2 and their experimental limit.}
\label{table:wilson}
\end{table}

For the contribution to $Br(B\to X_s \gamma)$ we on average get for the dominant Wilson coefficient,
\begin{equation}
 \left | \frac {C_7'}{C_7^{SM}} \right |  \, \approx \, 0.2
\end{equation}
which is much smaller than the experimentally allowed value $\sim 1.4$, see e.g.~\cite{2siteskaustubh}. 
As in the anarchic case the contribution to $C_7$ is small and can be neglected. With  CP protection 
all the points in this model are allowed by the data with a compositeness scale 
of 3 TeV. Flavor bounds would allow even lower scales of compositeness. 
\begin{figure}[h]
\begin{center}
\includegraphics[scale=1.0]{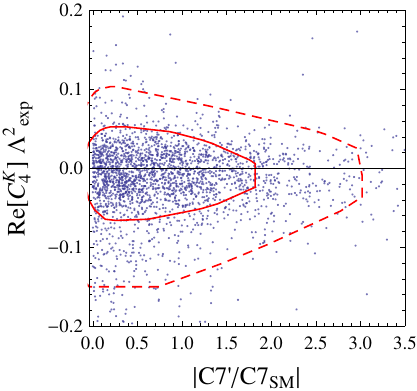}\hspace{.7cm}
\includegraphics[scale=1.0]{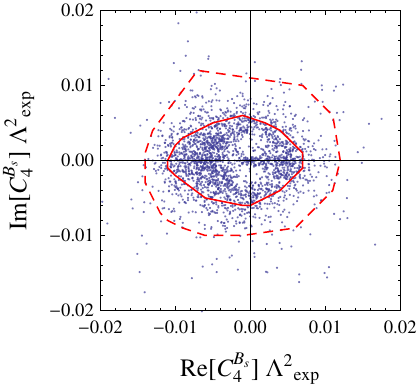}
\caption{Scatter plot for the second example with $g_\rho=6$. On the left $|C_7'/C_7^{SM}|$ and real part $C_4^K$. 
On the right we show the real and  imaginary parts of $C_4^{B_s}$. The solid (dashed) red lines enclose 68\% (95\%) of the points.}
\label{fig:c42}
\end{center}
\end{figure}

As a second example we consider stronger couplings for the composite sector. We use the following parameters,
\begin{center}
\begin{tabular}{|c|c|c|c|c|c|c|c|c|}
\hline 
$m_\rho$~(GeV) & $g_{\rho}$ & $Y_u$ & $Y_d$ & $\sin \varphi_{t_L}$ & $\sin \varphi_{t_R}$ & $\sin \varphi_{b_L}$ & $\sin \varphi_{b_R}$ \\ \hline
$3000$ & $6$ & $6.1$ & $6.5$ & $0.25$ & $0.75$ & $0.0625$ & $0.035$ \\  \hline
\end{tabular}
\end{center}
The average size of the Wilson coefficients is very similar to the one of the previous example, see table \ref{table:wilson}.
This is expected since we scaled $g_\rho $ and $Y$  by the same amount. 
The average value of $C_7'/C_7^{SM}$ is now 0.9, close to the experimental bound.
In this case about 70\% of the points generated passes all bounds, see Fig. \ref{fig:c42}. 

The experimental signatures of this model would be quite similar to the anarchic CHM since
the light fermions are mostly elementary. The main difference is that the only sizable contribution to 
CP violation are in the third generation quarks which in particular could appear in the $B_{d,s}$ system.

\section{Conclusions}
\label{outlook}

In the first part of this paper we have shown how the hypothesis of MFV can be realized in  Composite Higgs Models. This allows 
to avoid  all flavor constraints which are problematic in the standard anarchic scenario.
The key assumption is that the composite sector respects a flavor symmetry which forbids flavor transitions.
The SM flavor structure is induced by the mixings with the elementary SM fermions, some of which are proportional to the SM Yukawas.
Realizing MFV demands that some chiralities of the light quarks have a large degree of compositeness, the size being determined by 
the one of third generation quarks. We have studied two scenarios with either composite  left-handed or right-handed quarks. 
For left-handed compositeness, precision measurements (in particular the hadronic width of the $Z$ and the unitarity of the CKM matrix)
impose severe constraints for a low scale of compositeness. These bounds however do not apply in the case of composite right-handed fermions. 
Here, the main constraint arises at present from compositeness bounds. Large or even full compositeness of right-handed 
quarks is allowed for $m_\rho\sim 3$ TeV and would produce spectacular signals at the LHC where composite states
could be produced with large cross-sections. We have outlined the experimental signatures of the models
leaving a detailed study of this exciting phenomenology to future work.

In our framework the MFV hypothesis is incompatible  with the dynamical generation of the flavor hierarchies. This could be
realized with a composite sector that is a strongly coupled CFT (or its Randall-Sundrum hologram). This motivates 
a different construction, studied in the second part of the paper, where the composite sector is CP trivial rather than flavor trivial. 
In this case, in order to reproduce the large CKM phase, CP violation must be induced from the  elementary mixings. 
We have shown that when CP violation originates 
from third generation quarks, the contribution to problematic CP odd observables associated to the light generations
is suppressed, allowing to pass all flavor and EDMs bounds. While not generic, this construction could be compatible with the dynamical 
generation of flavor at low scales, with a collider phenomenology similar to the one of anarchic scenarios.

\vspace{0.5cm} {\bf Acknowledgments:} 
We wish to thank Georgios Choudalakis, Leandro Da Rold, Roberto Franceschini, Gilad Perez Javi Serra and David Straub 
for useful discussions  about compositeness bounds. We are grateful to Riccardo Rattazzi for suggesting part of this project to us
and numerous discussions about composite Higgs models.

\appendix

\section{Relation to 5D Models}
\label{5dmodel}

In this appendix we show how flavor or CP symmetric CHM can be realized within 5D scenarios. 

For the case of a flavor invariant composite sector the construction is similar to the one in Ref.  
\cite{Cacciapaglia:2007fw}, where models with left-handed quark compositeness and conformal dynamics were considered. 
Note that, since there is no dynamical explanation of flavor, the 5D theory does not require AdS metric 
(corresponding to conformal symmetry of the 4D strong sector). Each SM field is the zero mode of 5D fields in 
a representation of $SU(2)_L\otimes SU(2)_R \otimes U(1)_X$. Because of the flavor symmetry, bulk fermions
associated to different flavors have equal bulk mass parameters so that, in absence of UV  kinetic terms, 
their zero modes and Kaluza-Klein states are identical.  We emphasize that our setup differs significantly from the ``shining flavor'' paradigm~\cite{rattazzizaffaroni}. 
This proposal requires the existence of a conformal sector which contains exactly two marginal scalar flavon operators transmitting
the  UV flavor breaking into the bulk and onto the IR brane. This can in principle lead to non-degenerate bulk masses
assuming that the composite sector has controlled and non-trivial flavor dynamics~\cite{Fitzpatrick:2008zza,shiningcsaki,perez}.

In our setup the flavor structure is introduced through kinetic terms (allowed by the SM gauge symmetry) localized on the 
UV boundary, and it is for this reason  external to the strong sector. In the 4D two site picture this translates 
into non-canonical kinetic terms for the elementary fields,
\begin{equation}
K_{ij}^\psi\,\, \bar{\psi}^i \, \ddslash{ D}\, \,\psi^j \qquad \qquad \psi=\, q_L, \, u_R, \,d_R
\end{equation}
and universal mixings $\lambda^0_{\psi}$ with the composites. $K^\psi$ is an hermitian matrix which is the sum of
the UV localized kinetic term and the bulk one proportional to the identity. We can diagonalize $K$ and rescale the fields to get to 
the canonical basis. We obtain the mixings,
\begin{equation}
\lambda_{\psi}=  \frac 1 {\sqrt{K^\psi_D}} U_\psi \lambda^0_{\psi}
\end{equation}
where $U_\psi$ is a unitary matrix. Choosing $K^\psi$ accordingly we can reproduce eqs. (\ref{eqcompleft},\ref{eqcompright}).

For the CP invariant composite sector, it works in a similar way. In this case the bulk and IR actions respect
CP so that we can choose all the parameters to be real. The formula above still holds where however 
$\lambda^0_{\psi}$ is a  diagonal and hierarchical matrix if flavor hierarchies are generated dynamically. 
CP phases appear only in the unitary matrix $U$ and become physical only in the presence of a non trivial $K^\psi$ matrix.
The phases are also physical when the SM zero mode lives in two different 5D fields.

\section{4-Fermi Operators}
\label{app4fermiop}

In this appendix we derive the structure of 4-Fermi operators obtained integrating out the strong sector dynamics.
We will consider in particular the operators mediated by gluon resonances and flavor gauge bosons 
which we both expect to exist in our scenario. In principle their masses could be different and the 
flavor gauge bosons could be heavier since they do not play a role in stabilizing the electro-weak scale
but we will assume conservatively that they are at $m_\rho$.

We begin with gluon resonances. At the level of the strong sector we obtain the operator,
\begin{equation}
\frac {g_\rho^2}{2\, m_\rho^2}\,\left( \bar{Q}_\alpha^i \gamma^\mu T^a_{\alpha\beta} Q_\beta^i\, \bar{Q}_\gamma^j \gamma_\mu T^a_{\gamma\delta} Q_\delta^j\right)
\end{equation}
where greek indexes are color and latin include flavor. Each fermion is here Dirac and each bilinear is summed over
different species. By using the property of $SU(N)$ fundamental matrices,
\begin{equation}
T^a_{ij} T^a_{lm}=\frac 1 2\left( \delta_{im} \delta_{jl}-\frac 1 N \delta_{ij}\delta_{lm}\right) 
\end{equation} 
we obtain,
\begin{equation}
\frac {g_\rho^2}{4\, m_\rho^2}\, \left(\bar{Q}_\alpha^i \gamma^\mu  Q_\beta^i \bar{Q}_\beta^j \gamma_\mu  Q_\alpha^j-\frac 1 3 \bar{Q}_\alpha^i \gamma^\mu  Q_\alpha^i \bar{Q}_\beta^j \gamma_\mu  Q_\beta^j\right)
\label{gluonres}
\end{equation}
To obtain the low energy SM operators we dress these operators with the mixings.  If for example 
left-handed quarks are partially composite (first scenario) one obtains the LL operator
\begin{equation}
\frac {g_\rho^2}{4\, m_\rho^2} \sin^4\varphi_{q_L}\, \left(\bar{q}_{L\alpha}^i \gamma^\mu  q_{L\beta}^i \bar{q}_{L\beta}^j \gamma_\mu  q_{L\alpha}^j-\frac 1 3 \bar{q}_{L\alpha}^i \gamma^\mu  q_{L\alpha}^i \bar{q}_{L\beta}^j \gamma_\mu  q_{L\beta}^j\right)
\label{gluonapp}
\end{equation}
For LR operators we find that the  coefficient scales as,
\begin{equation}
\frac  {g_\rho^2}{4 m_\rho^2} \sin^2 \varphi_{q_{Li}}\sin^2 \varphi_{q_{Rj}}
\end{equation}
Similarly to flavor violating LR operators in eq. (\ref{LRoperator}), this coefficient can be expressed in terms of the quark masses. 
As a consequence only the operator containing two light quarks and  two top will have a sizable coefficient $\sim 1/m_\rho^2$.

Consider now flavor gauge bosons. For simplicity we focus on $U(3)$ flavor gauge bosons but very 
similar results hold with $U(3)_U\otimes U(3)_D$. Integrating them out we generate the operator,
\begin{equation}
\frac {g_\rho^2}{2\, m_\rho^2}\, \left(\bar{Q}_\alpha^i \gamma^\mu T^a_{ij} Q_\alpha^j \bar{Q}_\beta^k \gamma_\mu T^a_{kl} Q_\delta^l\right)
\end{equation}
By Fierzing this gives,
\begin{equation}
\frac {g_\rho^2}{4\, m_\rho^2}\, \left(\bar{Q}_\alpha^i \gamma^\mu Q_\alpha^j \bar{Q}_\beta^j \gamma_\mu Q_\delta^i\right)
\end{equation}
After mixing to the elementary quarks this contains (in the first scenario),
\begin{equation}
\frac {g_\rho^2}{4\, m_\rho^2}\, \sin^4 \varphi_{q_L}\,\left(\bar{q}_{L\alpha}^i \gamma^\mu  q_{L\beta}^i \bar{q}_{L\beta}^j \gamma_\mu  q_{L\alpha}^j\right)
\label{flavorapp}
\end{equation}
which is valid before rotation to the physical mass basis.
Contrary to eq. (\ref{gluonres}), this operator mediates flavor violating effects between the up and down sector. 
These effects are however suppressed with respect to the SM tree level ones 
and do not  impose phenomenological constraints \cite{Cacciapaglia:2007fw}.

Concerning bounds from compositeness, as explained in section \ref{secompbound}, only operators containing
valence quarks contribute dominantly. These are obtained from the fragments of eqs. (\ref{gluonapp},\ref{flavorapp}) with 
only up or down quarks. Gluon resonances generate,
\begin{equation}
\frac {g_\rho^2}{6\, m_\rho^2}\, \sin^4 \varphi_{q_{L,R}}(\bar{q}_{L,R} \gamma^\mu q_{L,R})^2
\end{equation}
where $q=(u,d)$. The same operator is obtained from flavor gauge bosons with a coefficient 1/4.
Comparing with the experimental bound we then derive,
\begin{equation}
\sin^2 \varphi_{q_{L,R}}\lsim \frac {2} {g_\rho} \left(\frac{m_\rho}{3 \rm{TeV}}\right)
\end{equation}
assuming gluon and flavor resonances to have the same mass.
In the second scenario if only up right-handed quarks are composite we obtain a similar bound due because of the presence 
of only 2 up quarks in the proton. If on the other hand only right-handed down quarks were  composite the bound would be weaker.

\section{CKM matrix}
\label{CKMredefinition}

The CKM matrix in eq. (\ref{CKMmath}) is a general unitary matrix. In the standard 
form the CKM is parametrized as follows,
\begin{equation}
V^{CKM}=\left(\begin{array}{ccc}
c_{12}\,c_{13} & s_{12}c_{13} & s_{13} e^{-i \delta} \\
-s_{12} c_{23}-c_{12} s_{23} s_{13} e^{i \delta}& c_{12}c_{23}-s_{12}s_{23} s_{13} e^{i \delta} & s_{23} c_{13}\\
s_{12} s_{23}-c_{12} c_{23} s_{13} e^{i \delta} & -c_{12} s_{23}-s_{12} c_{23} s_{13} e^{i \delta} & c_{23} c_{13}
\end{array}\right)
\end{equation}
Starting from the matrix (\ref{CKMmath}) we reach the above parametrization redefining the
phases of the SM quarks,
\begin{eqnarray}
(u, c ,t)\to (e^{i \beta_1} u,\,e^{i \beta_2} c,\,e^{i \beta_3} t) \nonumber\\
(d, s, b)\to (e^{i \gamma_1} d,\,e^{i \gamma_2} s,\, b)
\end{eqnarray}
One finds,
\begin{eqnarray}
&&\beta_1= - {\rm arg}[V^{CKM}_{11}] - {\rm arg}[V^{CKM}_{12}]-{\rm arg}[V^{CKM}_{23}]- {\rm arg}[V^{CKM}_{33}]\nonumber \\
&&\beta_2= - {\rm arg}[V^{CKM}_{23}]\nonumber\\
&&\beta_3= - {\rm arg}[V^{CKM}_{33}]\nonumber\\
&&\gamma_1= {\rm arg}[V^{CKM}_{12}]+{\rm arg}[V^{CKM}_{23}]+ {\rm arg}[V^{CKM}_{33}]\nonumber\\
&&\gamma_2= {\rm arg}[V^{CKM}_{11}]+{\rm arg}[V^{CKM}_{23}]+ {\rm arg}[V^{CKM}_{33}]
\label{CKMphases}
\end{eqnarray}
We see that the condition that $\gamma_1- \gamma_2\ll 1$, necessary to suppress CP violating 
effects in the Kaon system, corresponds ${\rm arg}[V^{CKM}_{11}]- {\rm arg}[V^{CKM}_{12}]\ll1$. This automatically satisfied
if CP violation is induced from the top mixings. The same conclusion is obtained for the up sector since 
${\rm arg}[V^{CKM}_{33}]\sim 0$.


\begin{thebibliography}{99}


\bibitem{MFV1}

R.~S. Chivukula and H.~Georgi, ``{Composite Technicolor Standard Model},'' {\em
  Phys. Lett.} {\bf B188} (1987)
99; L.~J. Hall and L.~Randall, ``{Weak scale effective supersymmetry},'' {\em Phys.
  Rev. Lett.} {\bf 65} (1990)
2939--2942; E.~Gabrielli and G.~F. Giudice, ``{Supersymmetric corrections to epsilon prime
  / epsilon at the leading order in QCD and QED},'' {\em Nucl. Phys.} {\bf
  B433} (1995) 3--25;
A.~Ali and D.~London, ``{Profiles of the unitarity triangle and CP-violating
  phases in the standard model and supersymmetric theories},'' {\em Eur. Phys.
  J.} {\bf C9} (1999) 687--703;
A.~J. Buras, P.~Gambino, M.~Gorbahn, S.~Jager, and L.~Silvestrini, ``{Universal
  unitarity triangle and physics beyond the standard model},'' {\em Phys.
  Lett.} {\bf B500} (2001) 161--167.
  
\bibitem{MFV}
 G.~D'Ambrosio, G.~F.~Giudice, G.~Isidori and A.~Strumia,
  ``Minimal flavour violation: An effective field theory approach,''
  Nucl.\ Phys.\  B {\bf 645}, 155 (2002)
  [arXiv:hep-ph/0207036].  
   
\bibitem{georgikaplan}  
M. J. Dugan, H. Georgi and D. B. Kaplan, ÒAnatomy Of A Composite Higgs Model,Ó Nucl. Phys. B 254, 299 (1985).

\bibitem{RSreview}
C.~Csaki, ``TASI lectures on extra dimensions and branes, published in Boulder 2002, Particle physics and cosmology,'' arXiv:hep-ph/0404096; R.~ Sundrum, ``To the fifth dimension and back. (TASI 2004),' published in Boulder 2004, Physics in D, arXiv:hep-th/0508134; H.~Davoudiasl, S.~Gopalkrishna, E.~Ponton, J.~Santiago, ``Warped 5-Dimensional Models: Phenomenological Status and Experimental Prospects,'' 0908.1968 [hep-ph];
R.~Contino, ``The Higgs as a Composite Nambu-Goldstone Boson,''  [arXiv:1005.4269 [hep-ph]];
C.~Grojean,``New theories for the Fermi scale,'' [arXiv:0910.4976 [hep-ph]].
  
\bibitem{higgsless}
 C.~Csaki, C.~Grojean, H.~Murayama, L.~Pilo, J.~Terning,
  ``Gauge theories on an interval: Unitarity without a Higgs,''
  Phys.\ Rev.\  {\bf D69}, 055006 (2004);   C.~Csaki, C.~Grojean, L.~Pilo, J.~Terning,
  ``Towards a realistic model of Higgsless electroweak symmetry breaking,''
  Phys.\ Rev.\ Lett.\  {\bf 92}, 101802 (2004).
  [hep-ph/0308038].
  
\bibitem{flavorgeneration}
Y. Grossman and M. Neubert, ``Neutrino masses and mixings in non-factorizable geometry,'' Phys. Lett. B 474, 361 (2000) [hep-ph/9912408]; T. Gherghetta and A. Pomarol, ``Bulk fields and supersymmetry in a slice of AdS,'' Nucl. Phys. B 586, 141 (2000) [hep-ph/0003129]; S. J. Huber and Q. Shafi, ``Fermion masses, mixings and proton decay in a Randall- Sundrum model,'' Phys. Lett. B 498, 256 (2001) [hep-ph/0010195]; S. J. Huber, ``Flavour violation and warped geometry,'' Nucl. Phys. B 666, 269 (2003) [hep-ph/0303183].  
  
\bibitem{minimalcomposite}
  K.~Agashe, R.~Contino, A.~Pomarol,
  ``The Minimal composite Higgs model,''
  Nucl.\ Phys.\  {\bf B719}, 165-187 (2005).
  [hep-ph/0412089].   
  
\bibitem{weiler}
  C.~Csaki, A.~Falkowski and A.~Weiler,
  ``The Flavor of the Composite Pseudo-Goldstone Higgs,''
  JHEP {\bf 0809}, 008 (2008)
  [arXiv:0804.1954 [hep-ph]].  
  
\bibitem{rattazzizaffaroni}
  R.~Rattazzi, A.~Zaffaroni,
  ``Comments on the holographic picture of the Randall-Sundrum model,''
  JHEP {\bf 0104}, 021 (2001).
  [hep-th/0012248].    
  
\bibitem{Cacciapaglia:2007fw}
  G.~Cacciapaglia, C.~Csaki, J.~Galloway, G.~Marandella, J.~Terning and A.~Weiler,
  ``A GIM Mechanism from Extra Dimensions,''
  JHEP {\bf 0804}, 006 (2008)
  [arXiv:0709.1714 [hep-ph]].
   
\bibitem{2sitescontino}
  R.~Contino, T.~Kramer, M.~Son and R.~Sundrum,
  ``Warped/Composite Phenomenology Simplified,''
  JHEP {\bf 0705}, 074 (2007)
  [arXiv:hep-ph/0612180].
  
\bibitem{buras}
A.~J.~Buras, C.~Grojean, S.~Pokorski and R.~Ziegler,
``Fcnc Effects in a Minimal Theory of Fermion Masses,''
arXiv:1105.3725 [hep-ph].

\bibitem{silh}
  G.~F.~Giudice, C.~Grojean, A.~Pomarol {\it et al.},
  ``The Strongly-Interacting Light Higgs,''
  JHEP {\bf 0706}, 045 (2007).
  [hep-ph/0703164].
  
\bibitem{discrete}
G.~Panico, A.~Wulzer,
  ``The Discrete Composite Higgs Model,''
   [arXiv:1106.2719 [hep-ph]].  
   
\bibitem{4dcomposite}
  S.~De Curtis, M.~Redi, A.~Tesi,
  ``The 4D Composite Higgs,''
    [arXiv:1110.1613 [hep-ph]].   
  
\bibitem{protectionzbb}
  K.~Agashe, R.~Contino, L.~Da Rold {\it et al.},
  ``A Custodial symmetry for Zb anti-b,''
  Phys.\ Lett.\  {\bf B641}, 62-66 (2006).
  [hep-ph/0605341].  
  
\bibitem{utfit}
M.~Bona {\it et al.} [UTfit Collaboration],
``Model-Independent Constraints on Delta F=2 Operators and the Scale of New Physics,''
JHEP {\bf 0803} (2008) 049
[arXiv:0707.0636 [hep-ph]].


\bibitem{kkbar}
M.~Blanke, A.~J.~Buras, B.~Duling, S.~Gori and A.~Weiler,
``$\Delta$ F=2 Observables and Fine-Tuning in a Warped Extra Dimension with Custodial Protection,''
arXiv:0809.1073 [hep-ph]; M.~Bauer, S.~Casagrande, U.~Haisch and M.~Neubert,
``Flavor Physics in the Randall-Sundrum Model: II. Tree-Level Weak-Interaction Processes,''
JHEP {\bf 1009} (2010) 017
[arXiv:0912.1625 [hep-ph]].
    
\bibitem{cthdm}
  J.~Mrazek, A.~Pomarol, R.~Rattazzi, M.~Redi, J.~Serra and A.~Wulzer,
  ``The Other Natural Two Higgs Doublet Model,''
  arXiv:1105.5403 [hep-ph].  
  
\bibitem{custodians}
 R.~Contino, L.~Da Rold, A.~Pomarol,
  ``Light custodians in natural composite Higgs models,''
  Phys.\ Rev.\  {\bf D75}, 055014 (2007).
  [hep-ph/0612048].  
  
\bibitem{Csaki:2008eh}
C.~Csaki, A.~Falkowski and A.~Weiler,
``A Simple Flavor Protection for Rs,''
Phys.\ Rev.\ D {\bf 80} (2009) 016001
[arXiv:0806.3757 [hep-ph]].  

\bibitem{2siteskaustubh}
  K.~Agashe, A.~Azatov and L.~Zhu,
  ``Flavor Violation Tests of Warped/Composite SM in the Two-Site Approach,''
  Phys.\ Rev.\  D {\bf 79}, 056006 (2009)
  [arXiv:0810.1016 [hep-ph]].

\bibitem{Gedalia:2009ws}
  O.~Gedalia, G.~Isidori and G.~Perez,
  ``Combining Direct \& Indirect Kaon CP Violation to Constrain the Warped KK Scale,''
  Phys.\ Lett.\  B {\bf 682}, 200 (2009)
  [arXiv:0905.3264 [hep-ph]].
  
\bibitem{Baker:2006ts}
C.~A.~Baker {\it et al.},
``An Improved Experimental Limit on the Electric Dipole Moment of the Neutron,''
Phys.\ Rev.\ Lett.\ {\bf 97} (2006) 131801
[arXiv:hep-ex/0602020].
     
\bibitem{soni}
  K.~Agashe, G.~Perez, A.~Soni,
  ``Flavor structure of warped extra dimension models,''
  Phys.\ Rev.\  {\bf D71}, 016002 (2005).
  [hep-ph/0408134].
    
\bibitem{duccio}
  R.~Barbieri, G.~Isidori, D.~Pappadopulo,
  ``Composite fermions in Electroweak Symmetry Breaking,''
  JHEP {\bf 0902}, 029 (2009).
  [arXiv:0811.2888 [hep-ph]].  
  
\bibitem{perez}
  C.~Delaunay, O.~Gedalia, S.~J.~Lee {\it et al.},
  ``Ultra Visible Warped Model from Flavor Triviality and Improved Naturalness,''
  [arXiv:1007.0243 [hep-ph]];
C.~Delaunay, O.~Gedalia, S.~J.~Lee, G.~Perez and E.~Ponton,
``Extraordinary Phenomenology from Warped Flavor Triviality,''
arXiv:1101.2902 [hep-ph].
  
\bibitem{pdg}
 C. Amsler et al. [Particle Data Group], ÒReview of particle physics,Ó Phys. Lett. B 667 (2008) 1.  
  
\bibitem{peskinlane}
  E.~Eichten, K.~D.~Lane, M.~E.~Peskin,
  ``New Tests for Quark and Lepton Substructure,''
  Phys.\ Rev.\ Lett.\  {\bf 50}, 811-814 (1983).  
  
\bibitem{LHCcompositeness}  
 V.~Khachatryan {\it et al.}  [CMS Collaboration],
  ``Measurement of Dijet Angular Distributions and Search for Quark
  Compositeness in pp Collisions at $sqrt{s} = 7$ TeV,''
  Phys.\ Rev.\ Lett.\  {\bf 106} (2011) 201804
  [arXiv:1102.2020 [hep-ex]];  G.~Aad {\it et al.} [ ATLAS Collaboration ],
  ``Search for New Physics in Dijet Mass and Angular Distributions in pp Collisions at $\sqrt{s} = 7$ TeV Measured with the ATLAS Detector,''
  New J.\ Phys.\  {\bf 13}, 053044 (2011).
  [arXiv:1103.3864 [hep-ex]].
      
\bibitem{gillioz}
 M.~Gillioz,
  ``A Light composite Higgs boson facing electroweak precision tests,''
  Phys.\ Rev.\  {\bf D80}, 055003 (2009).
  [arXiv:0806.3450 [hep-ph]].

\bibitem{gluonresonance}
  K.~Agashe, A.~Belyaev, T.~Krupovnickas, G.~Perez, J.~Virzi,
  ``LHC Signals from Warped Extra Dimensions,''
  Phys.\ Rev.\  {\bf D77}, 015003 (2008),
  [hep-ph/0612015].

\bibitem{agashe}
  K.~Agashe, H.~Davoudiasl, S.~Gopalakrishna, T.~Han, G.~-Y.~Huang, G.~Perez, Z.~-G.~Si, A.~Soni,
  ``LHC Signals for Warped Electroweak Neutral Gauge Bosons,''
  Phys.\ Rev.\  {\bf D76}, 115015 (2007),   [arXiv:0709.0007 [hep-ph]];
B.~Lillie, L.~Randall and L.~T.~Wang,
``The Bulk Rs Kk-Gluon at the Lhc,''
JHEP {\bf 0709} (2007) 074
[arXiv:hep-ph/0701166];
K.~Agashe, S.~Gopalakrishna, T.~Han, G.~-Y.~Huang, A.~Soni,
  ``LHC Signals for Warped Electroweak Charged Gauge Bosons,''
  Phys.\ Rev.\  {\bf D80}, 075007 (2009),  [arXiv:0810.1497 [hep-ph]];
K.~Agashe, A.~Azatov, T.~Han, Y.~Li, Z.~-G.~Si, L.~Zhu,
  ``LHC Signals for Coset Electroweak Gauge Bosons in Warped/Composite PGB Higgs Models,''
  Phys.\ Rev.\  {\bf D81}, 096002 (2010),
  [arXiv:0911.0059 [hep-ph]].
  
\bibitem{singleproduction}
J. Mrazek and A. Wulzer, ÒA Strong Sector at the LHC: Top Partners in Same-Sign Dileptons,Ó Phys.\ Rev.\  {\bf D81}, 075006 (2010), 
0909.3977 [hep-ph]; 
J. A. Aguilar-Saavedra, ÒIdentifying top partners at LHC,Ó JHEP 0911, 030 (2009) [0907.3155 [hep-ph]];
A.~Atre, G.~Azuelos, M.~Carena, T.~Han, E.~Ozcan, J.~Santiago and G.~Unel,
``Model-Independent Searches for New Quarks at the Lhc,''
arXiv:1102.1987 [hep-ph].
  
 \bibitem{doubleproduction}
  R.~Contino, G.~Servant,
  ``Discovering the top partners at the LHC using same-sign dilepton final states,''
  JHEP {\bf 0806}, 026 (2008),
  [arXiv:0801.1679 [hep-ph]].
  
\bibitem{atlasdirectbound}
http://cdsweb.cern.ch/record/1356196

\bibitem{dijet163invpb}
http://cdsweb.cern.ch/record/1355704

\bibitem{madgraph}
https://server06.fynu.ucl.ac.be/projects/madgraph;
  J.~Alwall, P.~Demin, S.~de Visscher, R.~Frederix, M.~Herquet, F.~Maltoni, T.~Plehn, D.~L.~Rainwater {\it et al.},
  ``MadGraph/MadEvent v4: The New Web Generation,''
  JHEP {\bf 0709 } (2007)  028.
  [arXiv:0706.2334 [hep-ph]].

\bibitem{cteq}
  J.~Pumplin, D.~R.~Stump, J.~Huston, H.~L.~Lai, P.~M.~Nadolsky, W.~K.~Tung,
  ``New generation of parton distributions with uncertainties from global QCD analysis,''
  JHEP {\bf 0207 } (2002)  012.
  [arXiv:hep-ph/0201195 [hep-ph]].

\bibitem{nelson}
 A.~E.~Nelson,
  ``Naturally Weak CP Violation,''
  Phys.\ Lett.\  B {\bf 136}, 387 (1984);    S.~M.~Barr,
  ``Solving The Strong CP Problem Without The Peccei-Quinn Symmetry,''
  Phys.\ Rev.\ Lett.\  {\bf 53}, 329 (1984).
  
\bibitem{Fitzpatrick:2008zza}
A.~L.~Fitzpatrick, L.~Randall and G.~Perez,
``Flavor Anarchy in a Randall-Sundrum Model with 5D Minimal Flavor Violation And a Low Kaluza-Klein Scale,''
Phys.\ Rev.\ Lett.\ {\bf 100} (2008) 171604.
  
\bibitem{shiningcsaki}
C.~Csaki, G.~Perez, Z.~Surujon and A.~Weiler,
``Flavor Alignment via Shining in Rs,''
Phys.\ Rev.\ D {\bf 81} (2010) 075025
[arXiv:0907.0474 [hep-ph]].

\bibitem{Delaunay:2012cz} 
  C.~Delaunay, J.~F.~Kamenik, G.~Perez and L.~Randall,
  ``Charming CP Violation and Dipole Operators from RS Flavor Anarchy,''
  JHEP {\bf 1301}, 027 (2013)
  [arXiv:1207.0474 [hep-ph]].

\bibitem{straub}
 M.~Kšnig, M.~Neubert and D.~M.~Straub,
  ``Dipole operator constraints on composite Higgs models,''
  arXiv:1403.2756 [hep-ph].

\bibitem{Redi:2013eaa}   
M.~Redi, V.~Sanz, M.~de Vries and A.~Weiler,
  ``Strong Signatures of Right-Handed Compositeness,''
  JHEP {\bf 1308}, 008 (2013)
  [arXiv:1305.3818 [hep-ph]].  

\end{thebibliography}
\end{document}